# Marangoni flow at droplet interfaces: Three-dimensional solution and applications

M. Schmitt[1] and H. Stark[1]

*Institute of Theoretical Physics, Technical University Berlin - Hardenbergstraße 36, 10623 Berlin, Germany*

The Marangoni effect refers to fluid flow induced by a gradient in surface tension at a fluid-fluid interface. We determine the full three-dimensional Marangoni flow generated by a non-uniform surface tension profile at the interface of a self-propelled spherical emulsion droplet. For all flow fields inside, outside, and at the interface of the droplet, we give analytical formulas. We also calculate the droplet velocity vector $\mathbf{v}^D$, which describes the swimming kinematics of the droplet, and generalize the squirmer parameter $\beta$, which distinguishes between different swimmer types called neutral, pusher, or puller. In the second part of this paper, we present two illustrative examples, where the Marangoni effect is used in active emulsion droplets. First, we demonstrate how micelle adsorption can spontaneously break the isotropic symmetry of an initially surfactant-free emulsion droplet, which then performs directed motion. Second, we think about light-switchable surfactants and laser light to create a patch with a different surfactant type at the droplet interface. Depending on the setup such as the wavelength of the laser light and the surfactant type in the outer bulk fluid, one can either push droplets along unstable trajectories or pull them along straight or oscillatory trajectories regulated by specific parameters. We explore these cases for strongly absorbing and for transparent droplets.

## I. INTRODUCTION

Self-propelled particles swimming in fluids at low Reynolds number have recently gained a lot of attention[1–4]. Different methods to construct microswimmers exist. One idea is to generate a slip velocity field close to the swimmer's surface by different phoretic mechanisms that drags the particle forward. A typical example of an artificial swimmer is a nano- or micron-sized Janus colloid. It has two distinct faces that differ in their physical or chemical properties[5]. In the simplest realization, one face catalyzes a chemical reaction and the reactants set up a self-diffusiophoretic flow[6]. A combination of self-diffusio- and electrophoresis close to bimetallic Janus particles in a peroxide solution generates an electrochemical gradient to propel the swimmer[7,8]. Heating a Janus particle, where the thermal conductivity of both faces differs, generates a temperature gradient, in which the colloid moves. This effect is called thermophoresis[9]. Finally, in a binary solvent close to the critical point, the liquid around Janus colloids demixes locally, which also induces a self-diffusiophoretic flow[10].

Both, the individual swimming mechanisms of these Janus particles and other microswimmers, as well as their collective motion have evolved into very attractive research topics[4,11]. In fact, the study of collective motion in non-equilibrium systems has opened up a new field in statistical physics. Recent studies of collective motion also concentrate on the role of hydrodynamic flow fields[12–17].

An alternative realization of a self-propelled particle is an active emulsion droplet. Motivated by the experimental realization of such a swimming droplet[18] and our own work[19], we construct here first the full three-dimensional solution for the flow field inside and outside of the droplet. It is driven by a non-uniform surface tension profile at the droplet interface. Then, we present two illustrative examples, where it is necessary to use this full solution.

When two immiscible liquid phases are mixed, they form emulsion droplets, which are often stabilized by surfactants. Emulsion droplets can be prepared with a well-defined size. Since they can enclose very small quantities of matter down to single molecules, they are predestined as microreactors in which chemical or biological reactions take place. Therefore, they are an important building block in microfluidic devices[20,21]. Droplets are commonly divided into two classes: passive droplets, which move due to external forces, and active droplets, which swim autonomously without any external forces. This force-free swimming is a general signature of self-propelled particles[4].

Self-propelled active droplets in a bulk fluid have been studied in various experiments[18,22–26] including droplets coupled to a chemical wave[27]. Theoretical as well as numerical treatments include deformable and contractile droplets[28,29], droplets in a chemically reacting fluid[30], studies of the drift bifurcation of translational motion[31–35], droplets driven by nonlinear chemical kinetics[36], and a diffusion-advection-reaction equation for surfactant mixtures at the droplet interface[19].

In the first part of this paper we derive the flow field around an emulsion droplet initiated by a non-uniform surface tension at the droplet interface. This phenomenon is known as Marangoni effect[37]. In the proximity of the droplet interface, Marangoni flow is directed towards increasing surface tension. So far, there have been detailed studies of the flow field around active droplets, where the surface tension is axisymmet-ric $\sigma = \sigma(\theta)$[38,39]. Formulas of the non-axisymmetric case have been mentioned in an extensive study of the rheology of emulsion drops and have been used to explain cross-streamline migration of emulsion droplets in Poiseuille flow[40–42]. Here, we present a detailed derivation and illustration of the full flow field for an arbitrary surface tension profile $\sigma(\theta, \varphi)$ at the droplet interface. We provide formulas for the flow fields inside and outside of the droplet, the droplet velocity vector $\mathbf{v}^D$, as well as for the squirmer parameter $\beta$, which determines whether a droplet is a pusher or a puller.

In the second part of this paper, we apply the presented formulas to two illustrative examples. There are various causes for a non-uniform surface tension field $\sigma(\theta, \varphi)$. A surfactant lowers the surface tension by accumulating at an interface. Thus the simplest way to generate Marangoni flow is a non-uniform distribution of a surfactant within an interface. In our first, simple example, we consider an initially "clean" or surfactant free droplet[43] immersed in a fluid, which is enriched by micelles, i.e., aggregates of surfactant molecules. When a micelle adsorbs somewhere at the droplet interface, Maragoni flow is induced and propels the droplet in the direction of the adsorption site. Now, the resulting outer fluid flow preferentially advects other micelles towards the existing adsorption site. This mechanism can spontaneously break the isotropic symmetry of the droplet, which then moves persistently in one direction, if the mean adsorption rate of the micelles is sufficiently large. Micelles have been shown to be crucial in the dynamics of active water as well as liquid crystal droplets[25,26]. While we do not attempt to unravel the detailed mechanism for activity in these examples, we present here a simple idea how micelle adsorption generates directed motion.

In the second example, we use a non-uniform mixture of two surfactant types to induce Marangoni flow. Such a mixture can be created by a chemical reaction[18]. Here, we illustrate a different mechanism. Light-switchable surfactants exist which change their conformation under illumination with light[44]. So, by shining laser light onto a droplet covered by light-switchable surfactants[44], one locally generates a spot of different surfactant molecules. Depending on the surfactant type in the bulk fluid and the wavelength of the laser light, the emulsion droplet is either pushed by the laser beam or pulled towards it. The first situation is unstable and the droplet moves away from the beam and then stops. In the second situation, the droplet moves on a straight trajectory along the beam. With decreasing relaxation rate towards the surfactant in bulk a Hopf bifurcation occurs and the droplet also oscillates about the beam axis. We explore these cases for strongly absorbing and for transparent droplets.

The article is organized as follows. In Sec. II we derive the flow fields inside and outside an emulsion droplet induced by a non-uniform surface tension profile. The flow fields depend on the droplet velocity vector $\mathbf{v}^D$, which we evaluate and discuss in Sec. III. Section IV discusses characteristics of the flow field and introduces the squirmer parameter in order to classify active droplets as pushers or pullers. The following two sections contain the illustrative examples. Section V demonstrates how micelle adsorption spontaneously breaks the isotropic droplet symmetry and induces directed propulsion. Finally, in Sec. VI we introduce and discuss the emulsion droplet covered by a light-switchable surfactant. The article concludes in Sec. VII.

## II. VELOCITY FIELD OF A FORCE-FREE ACTIVE EMULSION DROPLET

In the following, we consider a droplet of radius $R$ with viscosity $\hat{\eta}$ of the inside liquid immersed in an unbounded bulk fluid with viscosity $\eta$. At low Reynolds number we have to solve the creeping flow or Stokes equation to determine both the velocity field $\mathbf{u}(\mathbf{r})$ outside the droplet ($r > R$) and the field $\hat{\mathbf{u}}(\mathbf{r})$ inside the droplet ($r < R$). Solving the problem needs two steps[45].

At first we solve the Stokes equation for a droplet, which is fixed in space, with a given inhomogeneous surface tension $\sigma$ at the interface. The resulting flow field of this "pumping problem" will be called $\mathbf{w}$. Secondly, we derive the flow field $\mathbf{v}$ of a passive droplet swimming with a prescribed velocity $\mathbf{v}^D$, the so-called Hadamard Rybczynski solution[46]. The complete flow field of the swimming droplet is then given by the superposition of both flow fields: $\mathbf{u} = \mathbf{v} + \mathbf{w}$. This approach ensures that the swimming droplet is a force-free swimmer[4]. The droplet velocity vector $\mathbf{v}^D$ is calculated by means of the Lorentz reciprocal theorem for Stokes flow in Sec. III.

### A. Pumping active droplet

In this section, we fix the active emulsion droplet in space and analyze the velocity fields outside ($\mathbf{w}$) and inside ($\hat{\mathbf{w}}$) of the droplet generated by the inhomogeneous surface tension at the fluid interface. We start with the boundary conditions formulated in spherical coordinates in the droplet frame of reference:

$$\mathbf{w} = \mathbf{0}, \quad r \to \infty, \tag{1}$$

$$w_r = \hat{w}_r = 0, \quad r = R, \tag{2}$$

$$\mathbf{w} = \hat{\mathbf{w}}, \quad r = R, \tag{3}$$

$$\nabla_s \sigma = \mathbf{P}_s(\hat{\mathbf{T}} - \mathbf{T})\mathbf{e}_r, \quad r = R, \tag{4}$$

where $r$ is the distance from the droplet center and $R$ the droplet radius. These conditions assure that the droplet is fixed in space (1), has an impenetrable interface (2), and the tangential velocity at the interface is continuous (3). Condition (4) states that a gradient in surface tension $\sigma$ at the interface has to be balanced by a jump in the fluid shear stresses. This gradient in surface tension induces the Marangoni flow close to the interface. Here, we introduce the surface projector $\mathbf{P}_s = \mathbf{1} - \mathbf{n} \otimes \mathbf{n}$ with surface normal $\mathbf{n} = \mathbf{e}_r$. Correspondingly we use the notation $\nabla_s = \mathbf{P}_s \nabla$ for the surface gradient, where $\nabla$ is the nabla operator. In addition, we assume the droplet to be undeformable, *i.e.,* of constant curvature $\nabla \cdot \mathbf{n} = 2/R$, and thus do not need to consider the normal stress balance at the interface. Hence, we are in the regime of small capillary number $Ca = R|\nabla_s \sigma|/|\sigma| \ll 1$,[33]. Finally, the viscous part of the Cauchy stress tensor of an incompressible Newtonian fluid with viscosity $\eta$ is given by $\mathbf{T} = \eta \left[ \nabla \otimes \mathbf{w} + (\nabla \otimes \mathbf{w})^T \right]$. In spherical coordinates we find the following two equations from condition (4) for the polar and azimuthal components of $\nabla_s \sigma$:

$$(\nabla_s \sigma)_\theta = \hat{\eta}(\partial_r \hat{w}_\theta - R^{-1} \hat{w}_\theta) - \eta(\partial_r w_\theta - R^{-1} w_\theta) \text{ ,} \quad (5a)$$
$$(\nabla_s \sigma)_\varphi = \hat{\eta}(\partial_r \hat{w}_\varphi - R^{-1} \hat{w}_\varphi) - \eta(\partial_r w_\varphi - R^{-1} w_\varphi) \quad (5b)$$

Fluid flow at the liquid-liquid interface is always driven by a gradient in $\sigma$, whereas pressure only acts in normal direction.

We now use the set of boundary conditions (1)-(3) and (5) to solve the Stokes equation $\eta \nabla^2 \mathbf{w} - \nabla p = \mathbf{0}$ outside and inside the droplet. Due to the spherical symmetry of our problem and since pressure $p$ satisfies the Laplace equation, the following ansatz for the velocity and pressure fields outside the droplet are feasible according to Refs.[47,48]

$$\mathbf{w} = \sum_{l=0}^{\infty} \left[ \frac{2-l}{\eta l(4l-2)} r^2 \nabla p_l + \frac{1+l}{\eta l(2l-1)} p_l \mathbf{r} + \nabla \Phi_l + \nabla \times (\chi_l \mathbf{r}) \right], \quad (6)$$
$$p = \sum_{l=0}^{\infty} p_l .$$

Here, $\chi_l$, $\Phi_l$, and $p_l$ are irregular solid harmonics to assure $\mathbf{w} = 0$ as $r \to \infty$ according to Eq. (1):[48]

$$p_l = r^{-(l+1)} \sum_{m=-l}^{l} \alpha_l^m Y_l^m(\theta, \varphi) ,$$
$$\Phi_l = r^{-(l+1)} \sum_{m=-l}^{l} \beta_l^m Y_l^m(\theta, \varphi) ,$$

where we give the spherical harmonics $Y_l^m(\theta, \varphi)$ in appendix A. We already set the pseudoscalar $\chi_l$ in Eq. (6) to zero. We will express the flow field as a linear function in the surface tension, which is a scalar quantity. Due to the isotropic symmetry of the spherical droplet, a preferred direction does not exist and one cannot construct a term which contains the pseudoscalar $\chi_l$. Using the solid harmonics in Eq. (6) results in the following spherical components of the flow field $\mathbf{w}$ in Eq. (6),

$$w_r = \sum_{l=1}^{\infty} \sum_{m=-l}^{l} \left[ \frac{l+1}{(4l-2)\eta} \frac{\alpha_l^m}{r^l} Y_l^m - (l+1) \frac{\beta_l^m}{r^{l+2}} Y_l^m \right] ,$$
$$w_\theta = \sum_{l=1}^{\infty} \sum_{m=-l}^{l} \left[ \frac{2-l}{l(4l-2)\eta} \frac{\alpha_l^m}{r^l} \partial_\theta Y_l^m + \frac{\beta_l^m}{r^{l+2}} \partial_\theta Y_l^m \right] ,$$
$$w_\varphi = \sum_{l=1}^{\infty} \sum_{m=-l}^{l} \left[ \frac{im(2-l)}{l(4l-2)\eta} \frac{\alpha_l^m}{r^l} \frac{Y_l^m}{\sin\theta} + im \frac{\beta_l^m}{r^{l+2}} \frac{Y_l^m}{\sin\theta} \right] .$$

The coefficients $\alpha_l^m$ and $\beta_l^m$ will be determined in the following. Terms with $l = 0$ do not appear since the coefficients either vanish due to boundary conditions (1) and (2) ($\alpha_0^0 = \beta_0^0 = 0$).

The ansatz for the interior flow inside the droplet is obtained from Eq. (6) by replacing $l$ by $-(l+1)$ in the prefactor of each term:[47,48]

$$\hat{\mathbf{w}} = \sum_{l=0}^{\infty} \left[ \frac{(l+3)}{\hat{\eta}(l+1)(4l+6)} r^2 \nabla \hat{p}_l + \right.$$
$$\left. - \frac{l}{\hat{\eta}(l+1)(2l+3)} \hat{p}_l \mathbf{r} + \nabla \hat{\Phi}_l + \nabla \times (\hat{\chi}_l \mathbf{r}) \right], \quad (7)$$
$$\hat{p} = \sum_{l=0}^{\infty} \hat{p}_l ,$$

with regular solid harmonics, which do not diverge at $r = 0$,

$$\hat{p}_l = r^l \sum_{m=-l}^{l} \hat{\alpha}_l^m Y_l^m(\theta, \varphi) ,$$
$$\hat{\Phi}_l = r^l \sum_{m=-l}^{l} \hat{\beta}_l^m Y_l^m(\theta, \varphi) .$$

Again, we can set $\hat{\chi}_l = 0$. This results in the following spherical components of the flow field $\hat{\mathbf{w}}$ in Eq. (7),

$$\hat{w}_r = \sum_{l=1}^{\infty} \sum_{m=-l}^{l} \left[ \frac{l}{(4l+6)\hat{\eta}} r^{l+1} \hat{\alpha}_l^m Y_l^m + l r^{l-1} \hat{\beta}_l^m Y_l^m \right] ,$$
$$\hat{w}_\theta = \sum_{l=1}^{\infty} \sum_{m=-l}^{l} \left[ \frac{l+3}{(l+1)(4l+6)\hat{\eta}} r^{l+1} \hat{\alpha}_l^m \partial_\theta Y_l^m + r^{l-1} \hat{\beta}_l^m \partial_\theta Y_l^m \right] ,$$
$$\hat{w}_\varphi = \sum_{l=1}^{\infty} \sum_{m=-l}^{l} \left[ \frac{im(l+3)}{(l+1)(4l+6)\hat{\eta}} r^{l+1} \hat{\alpha}_l^m \frac{Y_l^m}{\sin\theta} + im r^{l-1} \hat{\beta}_l^m \frac{Y_l^m}{\sin\theta} \right] .$$

Terms with $l = 0$ are not relevant.

In the following, we successively evaluate the conditions (2), (3), and (5) to determine all the coefficients

$\alpha_l^m$, $\hat{\alpha}_l^m$ and $\beta_l^m$, $\hat{\beta}_l^m$. The condition of an impenetrable interface (2) connects $\alpha_l^m$ ($\hat{\alpha}_l^m$) with $\beta_l^m$ ($\hat{\beta}_l^m$):

$$\hat{\alpha}_l^m = \frac{-2\hat{\eta}(2l+3)}{R^2}\hat{\beta}_l^m \ , \ \alpha_l^m = \frac{-2\eta(1-2l)}{R^2}\beta_l^m \ .$$

We eliminate $\alpha_l^m$ and $\hat{\alpha}_l^m$ and use the condition $\hat{w}_\theta|_R = w_\theta|_R$ from Eq. (3) to relate the interior to the exterior coefficients:

$$\hat{\beta}_l^m = -\frac{l+1}{l}R^{-2l-1}\beta_l^m \ .$$

Evaluating $\hat{w}_\varphi|_R = w_\varphi|_R$ yields the same relations. In the next step we use the final condition (5) to match the jump in the shear stress to a given profile of the surface tension $\sigma$. From Eq. (5) and the coefficients determined above, we find:

$$(\nabla_s\sigma)_\theta = \sum_{l=1}^{\infty}\sum_{m=-l}^{l}\left[(\eta+\hat{\eta})\frac{4l+2}{l}\frac{\beta_l^m}{R^{l+3}}\partial_\theta Y_l^m\right] \ , \text{(8a)}$$

$$(\nabla_s\sigma)_\varphi = \sum_{l=1}^{\infty}\sum_{m=-l}^{l}\left[(\eta+\hat{\eta})\frac{4l+2}{l}\frac{im\beta_l^m}{R^{l+3}}\frac{Y_l^m}{\sin\theta}\right] \ . \text{(8b)}$$

We also expand the surface tension into spherical harmonics,

$$\sigma(\theta,\varphi) = \sum_{l=1}^{\infty}\sum_{m=-l}^{l}s_l^m Y_l^m(\theta,\varphi) \ , \tag{9}$$

with coefficients

$$s_l^m = \int\limits_0^{2\pi}\int\limits_0^{\pi}\sigma(\theta,\varphi)\overline{Y}_l^m(\theta,\varphi)\sin\theta\,\mathrm{d}\theta\,\mathrm{d}\varphi \ , \tag{10}$$

where $\overline{Y}_l^m$ is the complex conjugate of $Y_l^m$ (see appendix A). Thus, the l.h.s. of Eqs. (8) are given by:

$$(\nabla_s\sigma)_\theta = \frac{1}{R}\sum_{l=1}^{\infty}\sum_{m=-l}^{l}s_l^m\partial_\theta Y_l^m \ , \tag{11a}$$

$$(\nabla_s\sigma)_\varphi = \frac{1}{R}\sum_{l=1}^{\infty}\sum_{m=-l}^{l}ims_l^m\frac{Y_l^m}{\sin\theta} \ . \tag{11b}$$

Comparing Eqs. (8a) and (11a) or alternatively (8b) and (11b), we finally find

$$\beta_l^m = \frac{R^{l+2}}{\eta+\hat{\eta}}\frac{l}{4l+2}s_l^m \ .$$

This completes the derivation of the velocity field of an active droplet with a given surface tension profile $\sigma$, which is fixed in space.

The fluid flow at the interface is now easily calculated by inserting the coefficients $\beta_l^m$ into the ansatz for $\mathbf{w}$ and setting $r = R$:

$$w_\theta|_R = \frac{1}{\eta+\hat{\eta}}\sum_{l=1}^{\infty}\sum_{m=-l}^{l}\frac{s_l^m}{2l+1}\partial_\theta Y_l^m \ , \tag{12a}$$

$$w_\varphi|_R = \frac{1}{\eta+\hat{\eta}}\sum_{l=1}^{\infty}\sum_{m=-l}^{l}\frac{ims_l^m}{2l+1}\frac{Y_l^m}{\sin\theta} \ , \tag{12b}$$

with $s_l^m$ from Eq. (10).

Comparing the components of Eqs. (11) and (12) with each other, we realize that the expansion coefficients only differ by a factor $1/(2l+1)$. Thus, the fluid flow at the interface $\mathbf{w}|_R$ is basically equivalent to a smoothed gradient of the surface tension $\nabla_s\sigma$.

## B. Passive droplet

In this section we will calculate the velocity field $\mathbf{v}$ of the viscous flow around a passive sphere moving with a velocity $\mathbf{v}^D$. In the rest frame of the moving sphere boundary conditions (2) and (3) from the analysis of the fixed active droplet remain unchanged, while Eqs. (1) and (5) are replaced by

$$\mathbf{v} = -\mathbf{v}^D \ , \qquad r \to \infty \ , \tag{13}$$

$$\mathbf{0} = \mathbf{P}_s(\hat{\mathbf{T}} - \mathbf{T})\mathbf{e}_r \ , \qquad r = R \ . \tag{14}$$

The second condition is equivalent to $\nabla_s\sigma = 0$. It means that the fluid shear stress is continuous across the droplet interface and hence the droplet is passive. The procedure of calculating the flow field $\mathbf{v}$ is very similar to the case of the active droplet in Sec. II A. We outline it in the following.

We employ the same ansatz for the external ($\mathbf{v}$) and internal ($\hat{\mathbf{v}}$) droplet field, as we did for $\mathbf{w}$ and $\hat{\mathbf{w}}$ for the pumping active droplet in Eqs. (6) and (7). However, in order to satisfy boundary condition (13), we have to add the three spherical components of $-\mathbf{v}^D$,

$$-\mathbf{e}_r\cdot\mathbf{v}^D = -v_1^{-1}Y_1^{-1} - v_1^0Y_1^0 - v_1^1Y_1^1 \ ,$$
$$-\mathbf{e}_\theta\cdot\mathbf{v}^D = -v_1^{-1}\partial_\theta Y_1^{-1} - v_1^0\partial_\theta Y_1^0 - v_1^1\partial_\theta Y_1^1 \ ,$$
$$-\mathbf{e}_\varphi\cdot\mathbf{v}^D = -iv_1^1\frac{Y_1^1}{\sin\theta} + iv_1^{-1}\frac{Y_1^{-1}}{\sin\theta} \ ,$$

to the ansatz for $\mathbf{v}$, where we have introduced the coefficients $v_1^m$ for $\mathbf{v}^D$. This is equivalent to $\mathbf{v}^D$ in Cartesian representation:

$$\mathbf{v}^D = -\sqrt{\frac{3}{8\pi}}\begin{pmatrix} v_1^1 - v_1^{-1} \\ i\left(v_1^1 + v_1^{-1}\right) \\ -\sqrt{2}v_1^0 \end{pmatrix} \ . \tag{15}$$

From condition (2), we find:

$$\hat{\alpha}_l^m = \frac{-2\hat{\eta}(2l+3)}{R^2}\hat{\beta}_l^m \ ,$$
$$\alpha_l^m = \frac{-2\eta(1-2l)}{R^2}\left(\beta_l^m + \delta_{l,1}\frac{R^3}{2}v_l^m\right) \ ,$$

just as in the active case of Sec. II A. Boundary condition (3) relates $\hat{\beta}_l^m$ to $\beta_l^m$,

$$\hat{\beta}_l^m = -\frac{l+1}{l}R^{-2l-1}\beta_l^m + \frac{\delta_{l,1}}{2}v_l^m .$$

Finally, we use boundary condition (14) to derive:

$$\hat{\beta}_1^m = \frac{\eta}{2(\eta+\hat{\eta})}v_1^m ,$$

while all other coefficients with $l \geq 2$ vanish. So, we have related all coefficients to the components $v_1^m$ of $\mathbf{v}^D$.

We obtain the axially symmetric velocity field $\mathbf{v}$ of a passive droplet, the so called Hadamard-Rybczynski solution of a creeping droplet[46], which moves with velocity $\mathbf{v}_D$. In appendix B, where we give the complete velocity field $\mathbf{u}$ of the active droplet, one can read off the flow field $\mathbf{v}$ as the terms that contain $\mathbf{v}_D$. These terms either decay as $1/r$ or $1/r^3$. In particular, the velocity field at the interface is

$$v_\theta|_R = \frac{-\eta}{2(\eta+\hat{\eta})}\mathbf{e}_\theta \cdot \mathbf{v}^D , \quad (16a)$$

$$v_\varphi|_R = \frac{-\eta}{2(\eta+\hat{\eta})}\mathbf{e}_\varphi \cdot \mathbf{v}^D . \quad (16b)$$

We will calculate the droplet velocity $\mathbf{v}^D$ in Sec. III.

Note that, as $\hat{\eta} \to \infty$, one recovers the usual no-slip boundary condition of a rigid sphere.

### C. Complete solution

The complete flow field of the force-free swimming active droplet is the sum of both contributions, from the fixed active and the passive droplet. The velocity fields inside ($\hat{\mathbf{u}} = \hat{\mathbf{v}}+\hat{\mathbf{w}}$) and outside ($\mathbf{u} = \mathbf{v}+\mathbf{w}$) of the droplet are summarized in appendix B. The formulas are equivalent to the ones in Ref.[40–42]. The outside flow field is also presented in the lab frame. Here we mention the velocity field $\mathbf{u}|_R = \mathbf{w}|_R + \mathbf{v}|_R$ at the interface with $\mathbf{w}|_R$ taken from Eq. (12) and $\mathbf{v}|_R$ from Eq. (16):

$$u_\theta|_R = \frac{-\eta}{2(\eta+\hat{\eta})}\mathbf{e}_\theta \cdot \mathbf{v}^D$$
$$+\frac{1}{\eta+\hat{\eta}}\sum_{l=1}^{\infty}\sum_{m=-l}^{l}\frac{s_l^m}{2l+1}\partial_\theta Y_l^m , \quad (17a)$$

$$u_\varphi|_R = \frac{-\eta}{2(\eta+\hat{\eta})}\mathbf{e}_\varphi \cdot \mathbf{v}^D$$
$$+\frac{1}{\eta+\hat{\eta}}\sum_{l=1}^{\infty}\sum_{m=-l}^{l}\frac{ims_l^m}{2l+1}\frac{Y_l^m}{\sin\theta} . \quad (17b)$$

Before studying the axisymmetric limit, we investigate the role of viscosity. The most commonly studied droplet emulsions are either oil droplets in water or vice versa, where typical viscosities are $\eta_{\mathrm{water}} = 1\,\mathrm{mPa\,s}$ and $\eta_{\mathrm{oil}} = 36\,\mathrm{mPa\,s}$ [18,25]. We will show in the following section that

$\mathbf{v}^D \propto (2\eta+3\hat{\eta})^{-1}$. Using this result in Eqs. (17), we find that in the case $\eta \geq \hat{\eta}$ both $\mathbf{w}$ and $\mathbf{v}$ scale as $1/\eta$. In the opposite case, $\eta \ll \hat{\eta}$, the pumping solution scales as $\mathbf{w} \propto 1/\hat{\eta}$ while $\mathbf{v} \propto \eta/\hat{\eta}^2$. Hence, for an oil drop in water one can neglect $\mathbf{v}$, when calculating the velocity field (17) at the interface.

An axisymmetric surface tension $\sigma = \sigma(\theta)$, where only spherical harmonics with $m = 0$ contribute in Eqs. (9) and (17a), yields:

$$u_\theta|_R = \frac{\eta\sin\theta\,v_z^D}{2(\eta+\hat{\eta})} + \frac{1}{2(\eta+\hat{\eta})}\sum_{l=1}^{\infty}\left(\int_0^\pi \sigma P_l\sin\theta\mathrm{d}\theta\right)P_l^1 . \quad (18)$$

Here, $P_l(\cos\theta)$ are Legendre polynomials of degree $l$ and $P_l^1(\cos\theta) = \partial_\theta P_l(\cos\theta)$.

Levan et al. already solved the case of an axisymmetric swimming droplet[38], where the Stokes equation can be rephrased to a simpler fourth-order partial differential equation for a scalar stream function[48]. They found the flow field at the interface:

$$u_\theta|_R = \frac{\eta\sin\theta\,v_z^D}{2(\eta+\hat{\eta})}$$
$$+\frac{1}{2(\eta+\hat{\eta})}\sum_{l=2}^{\infty}l(l-1)\left(\int_0^\pi C_l^{-1/2}\partial_\theta\sigma\mathrm{d}\theta\right)\frac{C_l^{-1/2}}{\sin\theta} , \quad (19)$$

where $C_l^{-1/2}(\cos\theta)$ are Gegenbauer polynomials of order $l$ and degree $-1/2$. These are connected to Legendre polynomials by $\frac{d}{d\cos\theta}C_l^{-1/2} = -P_{l-1}$. A standard calculation, which uses the properties of Legendre and Gegenbauer polynomials, shows indeed that Eqs. (19) and (18) are equivalent.

The general solution for the surface flow, Eqs. (17), still contains the unknown droplet velocity vector $\mathbf{v}^D$. We will calculate $\mathbf{v}^D$ in Sect. III, by relating it to the non-uniform surface tension.

### III. DROPLET VELOCITY VECTOR

A central quantity in all studies of swimming droplets is the swimming speed $v^D$. Furthermore, for droplets without an axial symmetry, the swimming direction $\mathbf{e}$ is not obvious. Both together define the droplet velocity vector $\mathbf{v}^D = v^D\mathbf{e}$. Once this quantity is known, the flow field $\mathbf{u}|_R$ in (17) is completely determined.

In order to derive an expression for $\mathbf{v}^D$, we stress that an active particle is force-free[4]. Accordingly, the total hydrodynamic drag force $\mathbf{F} = \mathbf{F}_a + \mathbf{F}_p$, acting on the particle, has to vanish. Here, $\mathbf{F}_a$ and $\mathbf{F}_p$ are the drag forces of the active pumping droplet and the passive droplet, treated in Sec. II A and II B, respectively. The drag forces are given by $\mathbf{F}_a = -4\pi\nabla(r^3p_1|_a)$ and $\mathbf{F}_p = -4\pi\nabla(r^3p_1|_p)$, respectively, with solid harmonics $p_1|_a$ and $p_1|_p$ of the corresponding flow fields[48]. For the passive droplet one finds

$$\mathbf{F}_p = -6\pi\eta R \frac{2\eta + 3\hat{\eta}}{3\eta + 3\hat{\eta}} \mathbf{v}^D \ , \tag{20}$$

also known as the Hadamard and Rybczynski drag force of a droplet[46,49,50]. It reduces to the well known Stokes drag of a solid sphere for $\hat{\eta} \gg \eta$, whereas it predicts a reduced drag for droplets and bubbles, due to a finite slip velocity at the interface. The condition $\mathbf{F}_a + \mathbf{F}_p = 0$ gives the simple relation $\alpha_1^m|_a + \alpha_1^m|_p = 0$ between the coefficients determined in Secs. II A and II B and ultimately yields $v_1^m = -2s_1^m/(9\hat{\eta} + 6\eta)$. Here, $v_1^m$ are the coefficients of the velocity vector $\mathbf{v}^D$ introduced in Eq. (15). Thus, one finds:[41,51]

$$\mathbf{v}^D = \sqrt{\frac{1}{6\pi}} \frac{1}{2\eta + 3\hat{\eta}} \begin{pmatrix} s_1^1 - s_1^{-1} \\ i\left(s_1^1 + s_1^{-1}\right) \\ -\sqrt{2}s_1^0 \end{pmatrix} \ . \tag{21}$$

An equivalent relation writes $\mathbf{v}^D$ as the average of flow field $\mathbf{w}$ over the droplet surface, see appendix C. The droplet velocity vector is solely determined by the dipolar coefficients ($l = 1$) in the multipole expansion of the surface tension $\sigma$. It can be written as $\mathbf{v}^D = v^D \mathbf{e}$ with:

$$v^D = \frac{\sqrt{2\left(s_1^0\right)^2 - 4s_1^1 s_1^{-1}}}{\sqrt{6\pi}(2\eta + 3\hat{\eta})} \ , \tag{22a}$$

$$\mathbf{e} = \frac{1}{\sqrt{2\left(s_1^0\right)^2 - 4s_1^1 s_1^{-1}}} \begin{pmatrix} s_1^1 - s_1^{-1} \\ i\left(s_1^1 + s_1^{-1}\right) \\ -\sqrt{2}s_1^0 \end{pmatrix} \ . \tag{22b}$$

Next we derive an alternative formula for $\mathbf{v}^D$. Using the explicit expressions for the $s_1^m$ from Eq. (10) and the Cartesian components of the radial unit vector $\mathbf{e}_r$, we rewrite Eq. (21) as $\mathbf{v}^D = -[2\pi R^2(2\eta + 3\hat{\eta})]^{-1} \iint \sigma \mathbf{e}_r \mathrm{d}A$. Finally, extending $\sigma$ into the droplet with $\partial\sigma/\partial r = 0$ and applying Gauss's theorem, we obtain

$$\mathbf{v}^D = \frac{-1}{4\pi R(2\eta + 3\hat{\eta})} \iint \nabla_s \sigma \mathrm{d}A \ .$$

Thus, the droplet velocity vector $\mathbf{v}^D$ is simply given by the integral of the surface-tension gradient $\nabla_s \sigma$ over the whole droplet surface. By comparing this with the alternative formula for $\mathbf{v}^D$ in Eq. (C1), we realize: for calculating the droplet speed, the following equivalence holds:

$$\mathbf{w}|_R \,\hat{=}\, \frac{R}{3(\eta + \hat{\eta})} \nabla_s \sigma \ . \tag{23}$$

By using the $\hat{=}$ symbol, we stress that this equivalence is only valid in Eq. (C1) and not for the flow field $\mathbf{w}|_R$ in general. However, Eq. (23) illustrates that the surface flow is initiated by a gradient in the surface tension.

In the axisymmetric case ($m = 0$), $\mathbf{v}^D = -v^D \mathbf{e}_z$ points against the $z$ direction with

$$v^D = \frac{1}{2\eta + 3\hat{\eta}} \int_0^\pi \sigma \cos\theta \sin\theta \mathrm{d}\theta \ ,$$

which is equivalent to the swimming speed calculated by Levan et al.[38]. Note that the swimming velocity is independent of the droplet radius $R$.

Ansätze (6) and (7) for flow and pressure fields can also be used to treat droplets of non-spherical shape[48]. For the torque, which a droplet experiences from the surrounding fluid, one finds $\mathbf{M} = -8\pi\eta\nabla(r^3\chi_1)$,[48]. However, as explained in Sec. II A, the solid harmonic $\chi_1$ vanishes and thus $\mathbf{M} = 0$. Therefore, for a spherical droplet the angular velocity is zero, $\mathbf{\Omega} = 0$, and the swimming kinematics is completely determined by $\mathbf{v}^D$. Hence, there is no generalization of the Stokes drag torque $\mathbf{M} = -8\pi\eta R^3 \mathbf{\Omega}$ of a rigid particle to an emulsion droplet.

## IV. CHARACTERISTICS OF FLOW FIELD

In this section, we discuss some characteristics of the outside flow field $\mathbf{u}(\mathbf{r})$ fully presented in appendix B. Flow fields around an active particle can be written as a superposition of flow singularities[52,53]. The lowest singularity, the Stokeslet, is the flow field due to a point force $f\delta(\mathbf{r})\mathbf{a}$ pointing in direction $\mathbf{a}$. It decays as $\mathbf{u} \propto r^{-1}$ and is only present if external forces act on the particle. In our analysis we do not consider external forces.

The leading singularity of a force-free active droplet is the stresslet. An example is the force dipole constructed from two Stokeslets, which one obtains by taking the derivative of the Stokeslet along a given direction $\mathbf{b}$. The resulting flow field decays as $\mathbf{u} \propto r^{-2}$. In general, the stresslet corresponds to the symmetrized first moment of the force distribution on the particle surface. Thus, it is characterized by the symmetric tensor $\mathbf{S} = -\frac{2\pi}{3}\nabla \otimes \nabla(r^5 p_2)$ with solid harmonic $p_2$, which here comes from the pumping active droplet[54]. One obtains:

$$\mathbf{S} = \sqrt{\frac{6\pi}{5}} \frac{R^2\eta}{\eta + \hat{\eta}} \begin{pmatrix} s_2^{-2} - \sqrt{\frac{2}{3}}s_2^0 + s_2^2 & i(s_2^2 + s_2^{-2}) & s_2^{-1} - s_2^1 \\ i(s_2^2 + s_2^{-2}) & -s_2^{-2} - \sqrt{\frac{2}{3}}s_2^0 - s_2^2 & -i(s_2^{-1} + s_2^1) \\ s_2^{-1} - s_2^1 & -i(s_2^{-1} + s_2^1) & 2\sqrt{\frac{2}{3}}s_2^0 \end{pmatrix} \tag{24}$$

For instance, the flow field of two Stokeslets in direction $\mathbf{a} = \pm \mathbf{e}_x$, which are connected along $\mathbf{b} = \mathbf{e}_y$ is given by component $S_{xy}$. Clearly, only the coefficients $s_2^m$ account for the stresslet.

The singularities that account for a decay of the velocity field as $\mathbf{u} \propto r^{-3}$ are both the source dipole and the Stokes quadrupole. The source dipole is responsible for droplet propulsion and thus related to the droplet velocity vector $\mathbf{v}^D$ with the coefficients $s_1^m$ [see Eq. (21)]. Hence, these coefficients are always non-zero when the droplet is swimming. The Stokes quadrupole is, for example, a combination of two stresslets. It is related to the second moment of the force distribution, a tensor of rank three. As can be observed from appendix B, the coefficients $s_3^m$ account for the Stokes quadrupole. To summarize, the lowest-order decay of the flow field around a swimming active droplet is therefore either $\mathbf{u} \propto r^{-3}$ in the case of $\mathbf{S} = 0$ or $\mathbf{u} \propto r^{-2}$ if $\mathbf{S} \neq 0$.

### A. Generalized squirmer parameter

A useful parameter to quantify the type of a microswimmer driven by surface flow is the squirmer parameter $\beta$. It compares the stresslet strength to the source dipole. The squirmer is a classic model of an axisymmetric spherical microswimmer[55,56]. It has recently been generalized to the non-axisymmetric case[45]. The essential boundary condition of a squirmer is a prescribed flow field $\mathbf{w}|_R$ at the surface of a sphere. In contrast, the flow field $\mathbf{w}|_R$ of an active droplet is the result of a non-uniform surface tension at the fluid interface.

In the following, we calculate the squirmer parameter $\beta$ for an active droplet with arbitrary swimming direction as a function of the angular expansion coefficients $s_2^m$ of the surface tension. For an axisymmetric squirmer with surface flow velocity

$$u_\theta|_R = B_1 \sin\theta + \frac{B_2}{2}\sin 2\theta \ , \qquad (25)$$

one defines the squirmer parameter $\beta$ as[4,55–58]

$$\beta = B_2/|B_1| \ , \qquad (26)$$

where $2/3\,|B_1|$ is the swimming speed. When $\beta$ is positive, the surface flow is stronger in the front on the northern hemisphere and the flow around the squirmer is similar to the flow field initiated by a swimming algae such as *Chlamydomonas*. The swimmer is called a "puller" since it pulls itself through the fluid. Accordingly, a swimmer with $\beta < 0$ is called a "pusher" . For example, the bacterium *E. coli* swims by pushing fluid away from itself at the back by a rotating flagellum[4]. For $\beta \neq 0$, the flow field far away from the swimmer is dominated by the hydrodynamic stresslet or force dipole with $\mathbf{u} \propto r^{-2}$. However, in the case $\beta = 0$ ("neutral swimmer") the source dipole with $\mathbf{u} \propto r^{-3}$ dominates. One example for a neutral swimmer is the *Volvox* algae[4]. For $\beta \to \pm\infty$ the swimmer becomes a "shaker" that shakes the adjacent fluid but does not swim. Note that hydrodynamic interactions between swimmers as well as between swimmers and walls strongly depend on their type, *i.e.*, on the squirmer parameter $\beta$. Thus, $\beta$ is a key parameter in the study of individual swimmers as well as their collective dynamics[13,15,16,19].

For squirmers without axisymmetry but still swimming along the $z$-axis, Eq. (25) also contains terms depending on the azimuthal angle $\varphi$. In addition, a multipole expansion for the azimuthal velocity component $u_\varphi|_R$ has to be added. Still, the coefficient $B_1$ determines the swimming speed and $\beta$ the swimmer type since contributions from multipole terms with $m \neq 0$ vanish when averaging over $\varphi$.

So, we first determine the squirmer parameter for an axisymmetric droplet that swims in $z$ direction. Since $\beta$ is related to flow fields decaying like $1/r^2$, we only have to consider the velocity field $\mathbf{w}$ of the pumping active droplet. For the surface tension profile

$$\sigma = s_1^0 Y_1^0 + s_2^0 Y_2^0$$
$$= \sqrt{\frac{3}{4\pi}} s_1^0 \cos\theta + \sqrt{\frac{5}{4\pi}} s_2^0 \left( \frac{3}{2}\cos^2\theta - \frac{1}{2} \right) \ ,$$

where we have only included the relevant two leading modes, we find from Eq. (12a):

$$w_\theta|_R = \frac{-1}{\eta + \hat{\eta}} \left( \frac{s_1^0}{\sqrt{12\pi}}\sin\theta + \frac{1}{2}\frac{3s_2^0}{\sqrt{20\pi}}\sin 2\theta \right) \ .$$

Comparing with Eq. (25), we identify the squirmer parameter of the swimming axisymmetric droplet as the ratio

$$\beta = -\sqrt{\frac{27}{5}} \frac{s_2^0}{|s_1^0|} \ . \qquad (27)$$

In Eq. (D1) in appendix D, we relate all the angular coefficients $s_l^0$ to the squirmer coefficients $B_l$, which yields the same expression for $\beta$.

The ratio of stresslet tensor component $S_{zz} = \mathbf{e}_z \cdot \mathbf{S}\mathbf{e}_z$ and velocity $v^D$ is proportional to $\beta$ from Eq. (27). Similarly, when projecting the stresslet tensor $\mathbf{S}$ onto an arbitrary swimming direction $\mathbf{e}$, one averages over the azimuthal angle about $\mathbf{e}$. So the generalized squirmer parameter to characterize pushers and pullers becomes

$$\beta = -\frac{3}{4\pi R^2} \frac{\eta + \hat{\eta}}{(2\eta + 3\hat{\eta})\eta} \frac{\mathbf{e} \cdot \mathbf{S}\mathbf{e}}{v^D} \ . \qquad (28)$$

Here, we have set the prefactor such that $\beta$ agrees with Eq. (27) for $\mathbf{e} = \mathbf{e}_z$. In Eq. (E1) in appendix E, we give the concrete expression for $\beta$ in terms of $s_1^m$ and $s_2^m$. Note that in the non-axisymmetric case $\beta = 0$ does not mean that the stresslet is zero. For this all components of the stresslet tensor have to vanish. Only then one can conclude that a flow field with $\mathbf{u} \propto r^{-2}$ does not exist.

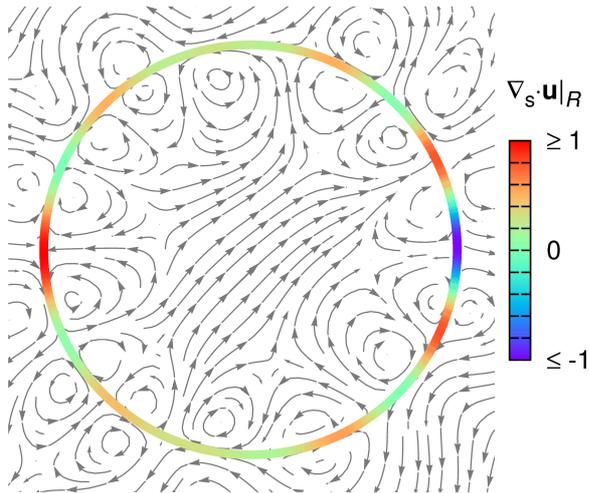

FIG. 1. Inside ($\hat{\mathbf{u}}$) and outside ($\mathbf{u}$) velocity field streamlines at a cross section through an emulsion droplet for a given surface tension $\sigma(\theta, \varphi)$. To draw the streamlines, the velocity vectors at the cross section are projected onto the cross section. Moreover, the surface divergence $\nabla_s \cdot \mathbf{u}|_R$ at the droplet interface is shown.

## B. Surface divergence

The solution of the Stokes equation for the flow field $\mathbf{u}$, which we presented in Sec. II C, fulfills the incompressibility condition $\nabla \cdot \mathbf{u} = 0$ everywhere, *i.e.*, also at the interface. However, this does not necessarily hold for the surface divergence $\nabla_s \cdot \mathbf{u}|_R$. Using $\nabla_s^2 Y_l^m = -l(l+1)R^{-2}Y_l^m$, one finds from Eqs. (17):

$$\nabla_s \cdot \mathbf{u}|_R = \frac{R^{-1}}{\eta + \hat{\eta}} \left[ \eta \mathbf{e}_r \cdot \mathbf{v}^D - \sum_{l=1}^{\infty} \sum_{m=-l}^{l} \frac{l(l+1)s_l^m}{2l+1} Y_l^m \right].$$ (29)

Thus any surface actuation $s_l^m \neq 0$ results in $\nabla_s \cdot \mathbf{u}|_R \neq 0$. In other words, surface divergence is a necessary condition for propulsion[25,59]. In fact, the surface divergence of the pumping field $\nabla_s \cdot \mathbf{w}|_R$, *i.e.*, the second term on the r.h.s. of Eq. (29), contains the expansion coefficients of $\sigma$ amplified by a prefactor $\mathcal{O}(l)$. Furthermore, comparing Eq. (29) with the radial components of the inside and outside velocity fields $\hat{u}_r$ and $u_r$ from (B1a) and (B1d), respectively, one finds the following. Regions at the interface with positive divergence, $\nabla_s \cdot \mathbf{u}|_R > 0$, are accompanied by radial flows $\hat{u}_r > 0$ and $u_r < 0$ towards the interface. On the other hand, regions with convergence, *i.e.*, negative divergence, $\nabla_s \cdot \mathbf{u}|_R < 0$, induce radial flows $\hat{u}_r < 0$ and $u_r > 0$ away from the interface. Figure 1 illustrates this. Depicted is the surface divergence $\nabla_s \cdot \mathbf{u}|_R$ along with the streamlines of $\hat{\mathbf{u}}$ and $\mathbf{u}$ at a cross section through an emulsion droplet with given surface tension field $\sigma(\theta, \varphi)$. In section V, we will build on this finding and demonstrate how micelle adsorption spontaneously breaks the isotropic symmetry of the droplet interface and thereby induces propulsion.

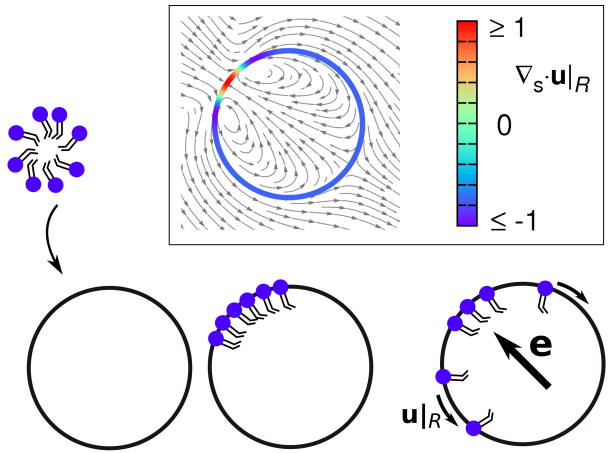

FIG. 2. Cartoon of a micelle adsorbing at the interface of an emulsion droplet. Marangoni flow $\mathbf{u}|_R$ spreads the surfactants over the interface and propels the droplet in direction $\mathbf{e}$ towards the adsorption site. Inset: Flow field and color-coded surface divergence $\nabla_s \cdot \mathbf{u}|_R$ shortly after a micelle has adsorbed at the droplet interface. Same representation as in Fig. 1.

## V. SPONTANEOUS SYMMETRY BREAKING BY MICELLE ADSORPTION

For the remainder of this paper, we will discuss possible applications for Marangoni flow initiated at the interface of an emulsion droplet. In this section we present a model of a droplet which performs directed motion by adsorbing micelles, *i.e.*, spherical aggregates of surfactant molecules. We consider a spherical oil droplet in water, the surface of which, initially, is hardly covered by surfactant molecules. Due to the small ratio of viscosities outside and inside the droplet, $\eta \ll \hat{\eta}$, we neglect the passive part $\mathbf{v}$ of the velocity field and set $\mathbf{u} = \mathbf{w}$, as pointed out in Sec. II C. The surrounding water phase is homogeneously enriched with micelles formed by surfactant molecules. In the following, we explain how this setup can lead to a persistent swimming motion of the droplet. Once one of the micelles with radius $R_M$ hits the droplet interface, the surfactants will adsorb at the droplet interface with a probability $p$ and cover a circular region of area $4\pi R_M^2$, as illustrated in Fig. 2. Thus, at the adsorption site surface tension is lower compared to the surrounding surfactant-free interface. The resulting Marangoni flow is directed away from the adsorption site and therefore spreads the surfactants over the droplet interface. The interfacial Marangoni flow $\mathbf{u}|_R = \mathbf{w}|_R$ induces a displacement of the droplet in the direction of the adsorption site with a velocity given in Eq. (C1). Furthermore, the flow is accompanied by a positive surface divergence $\nabla_s \cdot \mathbf{u}|_R > 0$ and inward radial flow $u_r < 0$ at the front of the droplet, as discussed in Sec. IV B. The flow field initiated by an adsorbed micelle is illustrated in the inset of Fig. 2. Now, the radial flow towards the interface advects additional micelles and thereby increases the rate with which surfactants adsorb at the front of the droplet. Following this train of thought, we expect the droplet to eventually develop a spot with increased surfactant coverage thereby breaking the isotropic symmetry of the interface. As a result, the droplet performs directed motion that comes to an end when the interface is fully covered by surfactants.

Note that this model starts with the assumption that the interface of the emulsion droplet initially is almost surfactant free. Such systems exist and Ref.[43] summarizes recent advances on realizing surfactant-free emulsion droplets. Furthermore, we do not take into account the detailed kinetics of the micelle adsorption[60]. We rather assume that when micelles adsorb at the interface they simply spread their surfactant molecules.

### A. Diffusion-advection equation

We propose a simple model for the surfactant dynamics at the droplet interface. The surfactant concentration $\Gamma$ obeys a diffusion-advection equation with additional source term,

$$\partial_t \Gamma = -\nabla_s \cdot (-D_s \nabla_s \Gamma + \Gamma \mathbf{u}|_R) + q \ . \tag{30}$$

The two terms in brackets describe transport of surfactants due to diffusion and advection induced by Marangoni flow, respectively, and $D_s$ is the diffusion constant within the interface. The third term on the r.h.s. of Eq. (30) represents the bulk current of micelles hitting the droplet interface, where they are ultimately adsorbed with a mean rate $1/\tau_{ads}$ and $\tau_{ads}$ is mean adsorption time.

The Marangoni flow $\mathbf{u}|_R$ at the droplet interface is generated by a concentration dependent surface tension, which we assume to be linear in $\Gamma$, for simplicity:

$$\sigma(\Gamma) = \sigma_0 - \sigma_s \Gamma \ , \tag{31}$$

where $\sigma_0 > \sigma_s > 0$. Here $\sigma_0$ is the surface tension of the clean or surfactant free droplet ($\Gamma \ll 1$) and $\sigma_0 - \sigma_s$ the surface tension of a droplet, which is fully covered by surfactants ($\Gamma = 1$). Thus, for a given surfactant density $\Gamma(\theta, \varphi)$, Eq. (31) yields the field of surface tension, which is expanded into spherical harmonics with coefficients $s_l^m$ according to Eq. (10). Note that the equation of state (31) typically breaks down at large surface coverage. Here, we mainly focus on the early droplet dynamics, when the interface is only lightly covered with surfactants.

The micellar source term $q$ has two contributions. Micelles perform a random walk through the outer fluid and ultimately hit the droplet interface which acts as a sink for the micelles. This sets up a diffusive current towards the interface. More importantly, as soon as Marangoni flow is initiated, micelles are also advected towards the interface as quantified by the radial flow component $u_r$, which is connected to the surface divergence $\nabla_s \cdot \mathbf{u}|_R$ at the droplet interface, as outlined in Sec.

IV B and discussed in Ref.[61] in more detail. A rigorous study of the full 3D bulk diffusion-advection equation for the bulk concentration of micelles $c$ is beyond the scope of this paper. Instead, we proceed as follows. Comparing the time scale of micellar bulk diffusion $t_D = R^2/(6D) = \pi \eta R^2 R_M/(k_B T)$ to the time scale of bulk advection $t_A = R(\eta + \hat{\eta})/\sigma_s$, we find a Peclet number Pe $= t_D/t_A$ on the order of $10^3$. For the estimate we took $R_M = R/20 = 50$nm, $\eta = 1$mPa s, $\hat{\eta} = 36$mPa s, $\sigma_s = 1$mN/m, and room temperature. Due to the large Peclet number, we neglect bulk diffusion completely and consider advection only in the following. We view micelle adsorption to occur anywhere at the interface as a Poissonian process. Adsorption events are independent of each other and the mean adsorption time between the events is $\tau_{ads}$ as already mentioned. Micelles preferentially adsorb at positions on the interface with large $\nabla_s \cdot \mathbf{u}|_R > 0$, while they do not adsorb at locations with $\nabla_s \cdot \mathbf{u}|_R < 0$, where the radial flow is directed away from the droplet. We will explain the detailed implementation of the adsorption event in the following section.

We introduce a dimensionless form of the diffusion-advection equation (30) for the droplet interface rescaling lengths by droplet radius $R$ and times by diffusion time $\tau = R^2/D_s$:

$$\partial_t \Gamma = -\nabla_s \cdot (-\nabla_s \Gamma + M \Gamma \mathbf{u}|_R) + q \ . \tag{32}$$

As one relevant parameter, the so-called Marangoni number $M = \tau/\tau_A$ compares the typical advection time $\tau_A = R(\eta + \hat{\eta})/\sigma_s$ to $\tau$, where we used $\mathbf{u}|_R = \sigma_s/(\eta + \hat{\eta})$ to estimate the Marangoni flow. Furthermore, the reduced adsorption rate becomes $\kappa = \tau/\tau_{ads}$, which is the most important parameter in this problem. Although we kept the same symbols, all quantities in Eq. (32), including $\Gamma$, $t$, $\nabla_s$, $\mathbf{u}|_R$, and $q$ are from now on dimensionless.

### B. Numerical solution

To solve Eq. (32) numerically, we used a finite-volume scheme on a spherical mesh[62]. Initially, at time $t = 0$, the droplet is free of surfactants, $\Gamma = 0$, and at rest. While diffusive and advective currents are implemented within the standard finite-volume algorithm, we model the micelle adsorption by a Poissonian process[63]. At each time step $\Delta t$ in the numerical scheme, we allow an adsorption event with probability $\kappa \Delta t$. If it is successful, the micelle is adsorbed with larger probability at positions were $\nabla_s \cdot \mathbf{u}|_R > 0$ is large. To implement this, we introduce the weight function

$$f(\theta, \varphi) = \frac{\nabla_s \cdot \mathbf{u}|_R}{\int \nabla_s \cdot \mathbf{u}|_R \mathrm{d}\Omega} \quad \text{for} \quad \nabla_s \cdot \mathbf{u}|_R > 0 \ . \tag{33}$$

Then the probability for micelle adsorption during time $\Delta t$ and within the solid angle element $\mathrm{d}\Omega$ at an angular

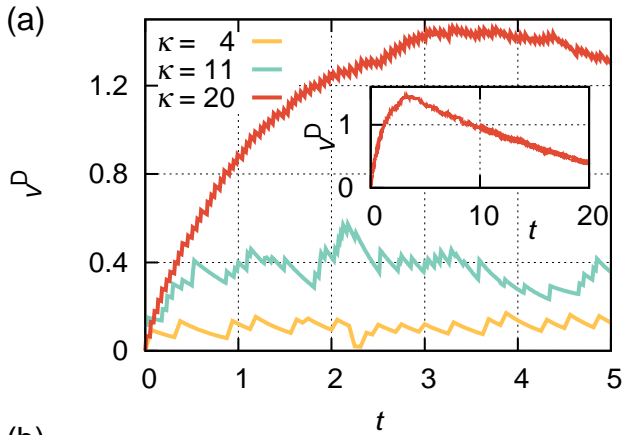

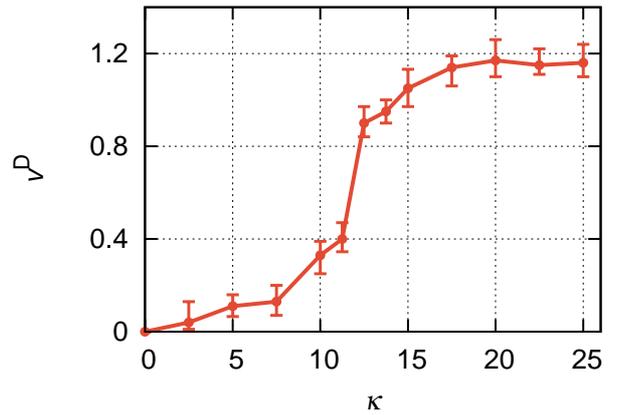

FIG. 4. Swimming speed $v^D$ plotted versus reduced adsorption rate $\kappa$. For each $\kappa$, speed $v^D$ is taken at time $t = 5$ and averaged over 60 simulation runs.

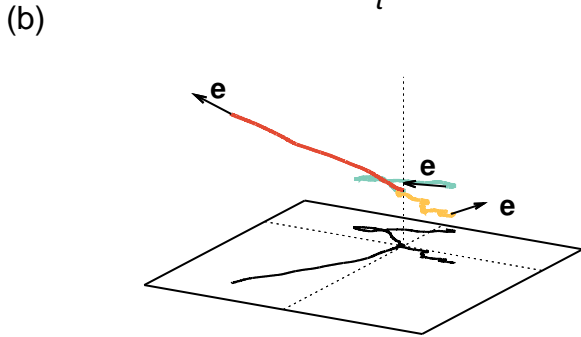

FIG. 3. (a) Swimming velocity $v^D$ plotted versus time $t$ for an emulsion droplet adsorbing surfactant micelles. The reduced adsorption rates are $\kappa = 4$, 11, and 20. Other parameters are $M = 1$ and $R/R_M = 20$. The inset shows the long-time limit for $\kappa = 20$. (b) Swimming trajectories from the same simulations as in (a) starting from $t = 0$ until the droplets stop when they are fully covered with surfactants.

position $(\theta, \varphi)$ becomes

$$p(\theta, \varphi, \Delta t)\mathrm{d}\Omega = \begin{cases} \kappa \Delta t f(\theta, \varphi)\mathrm{d}\Omega & \text{for } \nabla_s \cdot \mathbf{u}|_R > 0, \\ 0 & \text{for } \nabla_s \cdot \mathbf{u}|_R \leq 0. \end{cases}$$

After a micelle adsorption event at site $(\theta, \varphi)$ is determined, we set $\Gamma$ to one in a circular patch with radius $2R_M$ centered around $(\theta, \varphi)$. In addition, we assume that surfactants stay at the droplet interface once adsorbed.

In this setup Marangoni flow and diffusion current act in the same direction along $-\nabla_s\Gamma$. So, the Marangoni number is not the relevant parameter to initiate directed motion and we always set $M = 1$. However, by tuning the adsorption rate $\kappa$, the droplet starts to swim.

Figure 3 (a) shows the swimming speed $v^D$ of an emulsion droplet for three values of the reduced adsorption rate $\kappa = \tau/\tau_{\mathrm{ads}}$. For $\kappa = 4$, the mean adsorption time is too large. The surfactant patch from a first micellar impact has already spread over the whole interface by diffusion and advection when a second micelle hits the droplet interface at a different location. As a result, the droplet follows a random trajectory. Figure 3 (b) shows the corresponding swimming trajectory determined from $\mathbf{r}(t) = \mathbf{r}(0) + \int_0^t \mathrm{d}t' v^D(t')\mathbf{e}(t')$. Increasing

$\kappa$ to 11, increases the number of micelles, which adsorb per unit time, and the swimming speed becomes larger. The swimming trajectory is still irregular albeit with an increased persistence.

Finally, for $\kappa = 20$ mean adsorption time is significantly shorter than the characteristic diffusion time. Thus, when a second micelle is about to hit the droplet interface, surfactant concentration $\Gamma$ and surface divergence $\nabla_s \cdot \mathbf{u}|_R$ are still peaked at the impact of the previous micelle. Therefore, the probability of the following micelle to adsorb at the front of the droplet is increased compared to the back. This spontaneously breaks spherical symmetry. A defined swimming direction evolves and the droplet shows directed motion with swimming velocity $v^D$. This is confirmed by the swimming trajectory in Fig. 3 (b). As the droplet continues to swim, the difference in surfactant concentration at the adsorption site and the mean concentration at the interface decreases. As a consequence the Marangoni flow extenuates and $v^D$ decreases in time [see inset of Fig. 3 (a)]. Finally, when the interface is fully covered, $i.e.$, $\Gamma = 1$ on the whole interface, the droplet stops.

Figure 4 shows the onset of directed swimming by plotting swimming speed versus reduced adsorption rate $\kappa$. Due to amplification of micellar adsorption at a specific spot, the droplet switches at $\kappa \approx 12$ from slow, random motion with $v^D \approx 0.1 \ldots 0.4$ to fast and persistent motion with $v^D \approx 0.9 \ldots 1.2$.

In experiments, $\kappa = \tau/\tau_{\mathrm{ads}}$ can be tuned by adjusting the surfactant concentration $c_S$. We equate the micellar adsorption rate $\tau_{\mathrm{ads}}^{-1}$ with the flux $j \cdot 4\pi R^2$ of micelles from the bulk to the droplet interface. The current is advective, $j = c_M v_A$, with micelle concentration $c_M$ and velocity $v_A = R/t_A$. Using also $\tau = R^2/D_s$, one finds $\kappa = c_M \cdot 4\pi R^4/(D_s t_A)$. We assume that a micelle consists of $10^4$ surfactants and introduce the degree of micellization as the ratio $\gamma$ between micellized surfactants and all surfactants in the system, this yields $c_M = \gamma 10^{-4} c_S$. The ratio $\gamma$ changes strongly around the critical micelle concentration $c_{CMC}$. For example, for $c_S = 0.9 \cdot c_{CMC}$, i.e., slightly below $c_{CMC}$, one finds $\gamma \approx 5 \cdot 10^{-4}$,[64]. Together with estimates $D_s = 10^{-5} \mathrm{cm}^2/\mathrm{s}$,[65], $c_{CMC} = 1.5 \mathrm{mmol/l}$, and values for $R$ and $t_A$ from Sec. V A, we obtain $\kappa \approx 15$, thus around the onset of motion in Fig. 4. However, $\gamma \approx 10^{-4}$ means that micelle adsorption strongly competes with monomer adsorption, which is not contained in our model to keep it simple. Thus, to observe the onset of droplet motion in experiments, one has to increase $c_M$ by tuning the system closer to $c_{CMC}$ or even above.

Finally, we note that for increasing Marangoni number $M$, the patch of surfactants spreads faster due to advection and the crossover in Fig. 4 simply shifts towards larger $\kappa$.

## VI. LIGHT-INDUCED MARANGONI FLOW

Certain surfactants are known to be photosensitive[66–69]. For instance, surfactants based on azobenzene can undergo photoisomerization, where UV light (365 nm) transforms a *trans* to a *cis* configuration and blue light (450 nm) causes a transformation from *cis* to *trans*. During the *trans-cis* isomerization subunits within the molecule change their relative orientation. Naturally, a different molecular structure also affects the surface tension of a surfactant-covered interface. Experiments showed that surfactants in the *cis* state cause a higher surface tension compared to the ones in the *trans* state[67]. This effect has recently been used to generate Marangoni flow[44]. Therefore, we suggest two possible applications of the formulas presented in Sec. II. We first treat light-driven motion of a strongly absorbing, i.e., "dark", emulsion droplet and then discuss how the results alter in the case of a transparent droplet.

### A. Pushing an absorbing droplet with UV light

We think of an experiment where a spherical oil droplet of constant radius $R$ is placed in a water phase laden with *trans* surfactants. Initially, the emulsion droplet is in equilibrium with the exterior phase and thus completely covered with *trans* surfactants. This corresponds to times $t \gg \kappa^{-1}$ in Sec. V. A UV laser beam with cross-sectional radius $\rho < R$ is focused on the center of the droplet. It locally transforms surfactants at the interface into the *cis* state and thereby increases surface tension, see Fig. 5 (a).

Here we assume that the droplet oil phase completely absorbs the incident light beam. Accordingly, the laser beam does not reach the interface opposite to the illuminated side. A thinkable droplet phase is crude oil, which has a penetration depth of $\alpha \approx 100\mu m$ at wavelength $400\mathrm{nm}$[70]. On this length scale the droplet is still in the low Reynolds number regime and all findings of Secs. II - IV are valid. Alternatively, one may fabricate a "dark" droplet by enriching the oil phase with soot or black pigment. In Sec. VI C we study a transparent droplet.

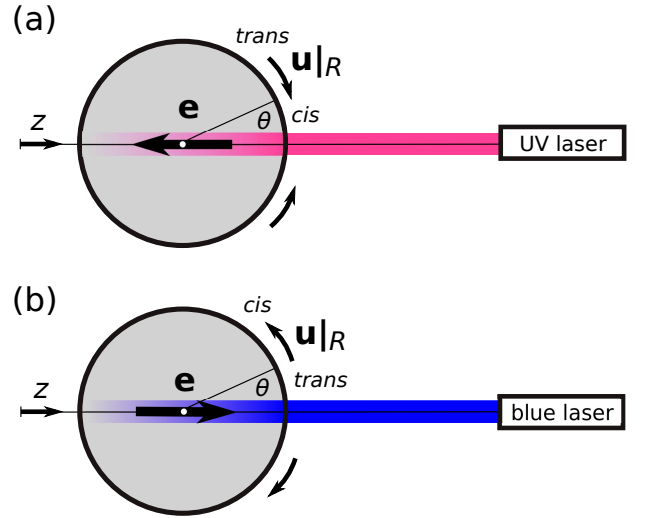

FIG. 5. (a) UV laser light is aimed at a strongly absorbing oil droplet in water. This increases the surface tension $\sigma$ locally at the droplet interface by transforming *trans* to *cis* surfactants. The resulting Marangoni flow $\mathbf{u}|_R$ is directed towards increasing surface tension and leads to propulsion in direction $\mathbf{e}$ away from the laser beam. (b) Blue laser light locally decreases the surface tension $\sigma$ at a droplet interface by transforming *cis* to *trans* surfactants. The resulting Marangoni flow $\mathbf{u}|_R$ leads to propulsion in direction $\mathbf{e}$ towards the laser beam.

The initiated Marangoni flow is oriented towards the laser beam and thus the droplet is propelled away from the laser beam, see Fig. 5 (a). Due to the advective current of surfactants towards the laser beam, the *cis* surfactants converge at the laser spot on the droplet interface and ultimately leave the interface. Fresh *trans* surfactants are adsorbed at the leading front of the droplet, i.e., at the side opposite to the laser beam.

#### 1. Diffusion-advection-reaction equation

In the following we review our theoretical approach to describe how the mixture of *trans* and *cis* molecules evolves in time, which then determines the dynamics of the flow field. More details can be found in Ref.[19]. We introduce the order parameter field $\phi(\theta, \varphi)$ with respective values $\phi = +1$ or $-1$ in regions where all surfactants are either in the *cis* or *trans* state, while in mixtures of both surfactants $\phi$ is in the range $-1 < \phi < 1$.

The dynamics of the order parameter $\phi$ at the droplet interface can be expressed by the diffusion-advection-reaction equation:

$$\partial_t \phi = -\nabla_s \cdot (\mathbf{j}_D + \phi \mathbf{u}|_R) - \tau_{eq}^{-1}(\phi - \phi_{eq}) , \quad (34)$$

with diffusive current $\mathbf{j}_D$ and advective Marangoni current $\phi \mathbf{u}|_R$. The source term couples the order parame-ter to the outer fluid laden with *trans* surfactants, *i.e.*, $\phi_{\mathrm{eq}} = -1$, by introducing a relaxation dynamics with timescale $\tau_{\mathrm{eq}}$.

To derive the diffusive current $\mathbf{j}_D$, we use a Flory-Huggins free energy density[19]

$$f(\phi) = \frac{k_B T}{\ell^2} \left[ \frac{1+\phi}{2} \ln \frac{1+\phi}{2} + \frac{1-\phi}{2} \ln \frac{1-\phi}{2} \right.$$

$$\left. - \frac{1}{4}(b_1 + b_2 + b_{12}) - \frac{\phi}{2}(b_1 - b_2) - \frac{\phi^2}{4}(b_1 + b_2 - b_{12}) \right], \tag{35}$$

where $\ell^2$ is the head area of a surfactant at the interface. We introduce dimensionless parameters $b_1$ and $b_2$ to characterize the respective interactions between either *cis* or *trans* surfactants and $b_{12}$ describes the interaction between the two types of surfactants. With the total free energy $F[\phi] = \int f(\phi) \, dA$ the diffusive current becomes:

$$\mathbf{j}_D = -\lambda \nabla_s \frac{\delta F}{\delta \phi} = -D_s \left[ \frac{1}{1-\phi^2} - \frac{1}{2}(b_1 + b_2 - b_{12}) \right] \nabla_s \phi \,, \tag{36}$$

where the Einstein relation $D_s = \lambda k_B T / \ell^2$ relates mobility $\lambda$ to the interfacial diffusion constant $D_s$. Note that, the condition $\mathbf{j}_D \propto -\nabla_s \phi$ is only fulfilled for a convex free energy with $f''(\phi) > 0$, *i.e.*, if $b_1 + b_2 - b_{12} < 2$. In the following we assume for simplicity $b_{12} = (b_1 + b_2)/2$.

In order to determine the Marangoni flow at the interface, we need an expression for the surface tension $\sigma$. From the free energy (35), we obtain the equation of state for the surface tension[19]:

$$\sigma(\phi) = \frac{k_B T}{\ell^2}(b_1 - b_2) \left[ \frac{3}{8} \frac{b_1 + b_2}{b_1 - b_2} + \frac{1}{2}\phi + \frac{1}{8} \frac{b_1 + b_2}{b_1 - b_2} \phi^2 \right]. \tag{37}$$

Hence, for a given order parameter profile $\phi(\theta, \varphi)$, Eq. (37) yields the field of surface tension, which is expanded into spherical harmonics with coefficients $s_l^m$ according to Eq. (10). Note that in contrast to the equation of state (31) of the single-surfactant model in Sec. V, expression (37) is nonlinear.

In order to make Eq. (34) dimensionless, we rescale lengths by droplet radius $R$ and time by the diffusion time scale $\tau = R^2/D_s$ and obtain

$$\partial_t \phi = -\nabla_s \cdot (\mathbf{j}_D + M\phi \mathbf{u}|_R) - \kappa(\phi - \phi_{\mathrm{eq}}) \,. \tag{38}$$

Here, we introduced the Marangoni number $M = \tau/\tau_A$, where $\tau_A = \ell^2 R(\eta + \hat{\eta})[(b_1 - b_2)k_B T]^{-1}$ is the advection time scale, and $\kappa = \tau/\tau_{\mathrm{eq}}$. All quantities of Eq. (38) are dimensionless. We numerically solve Eq. (38) on a spherical domain by the method of finite volumes as explained in detail in Ref.[71]. In all what follows, we set $b_1 = 2$ and $b_2 = 1$ as well as $M = 1$, *i.e.* $\tau = \tau_A$. Furthermore, we choose $\kappa = 1$ to illustrate the main behavior but also discuss the system's dynamics for different values of $\kappa$.

### 2. Stationary solution of pushed droplet

Initially, we set the order parameter $\phi$ to $-1$ on the whole interface. We then turn on the UV laser beam hitting the interface on a circular patch with radius $\rho = 0.2R$. In our numerical scheme, this is implemented by setting $\phi = 1$ in the area of exposure. Furthermore, to couple the droplet to the outer fluid laden by *trans* surfactants, we set $\phi_{\mathrm{eq}} = -1$. Figure 6, case (a) shows a typical stationary order parameter profile $\phi$, which results from the dynamics of Eq. (38). While $\phi$ exhibits a step-like function, the interfacial Marangoni flow $u_\theta|_R$, also illustrated in Fig. 6, case (a), spreads over the whole droplet interface. However, since the flow field is concentrated on the northern hemisphere and directed towards $\theta = 0$, the droplet is a pusher. This is confirmed by the formulas for the squirmer parameter from Sect. IV A, which yield $\beta = -2.8$. Increasing $\kappa$ enhances the coupling to $\phi_{\mathrm{eq}} = -1$ at the droplet front and the step in the order parameter profile $\phi$ becomes steeper, whereas the profile $\phi$ does not significantly depend on the Marangoni number $M$.

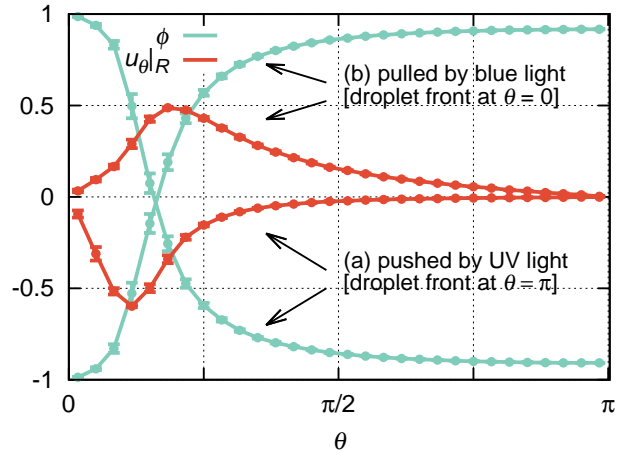

FIG. 6. Stationary solutions of the order parameter field $\phi$ and the flow field $u_\theta|_R$ for (a) the droplet which is pushed by UV light and (b) the droplet which is pulled by blue light. In both cases, the laser light hits the droplet interface at $\theta = 0$, compare Fig. 5. Order parameter $\phi = 1$ and $-1$ corresponds to pure *cis* and *trans* surfactants, respectively. Further parameters are $M = \kappa = 1$.

### 3. Pushing the droplet off-center

So far, the pushed droplet swims with a constant velocity $\mathbf{v}^D = -v^D \mathbf{e}_z$. Now, we introduce an offset $\Delta y$ of the UV laser beam from the center of the droplet, and study the impact on the droplet trajectory. Figure 7(a) illustrates the situation. Due to the offset $\Delta y$, the Marangoni flow $\mathbf{u}|_R$ pushes the droplet out of the laser beam. This increases the offset further and the orientation vector $\mathbf{e}$ tilts further away from the laser beam.

Figure 8 (a) shows the trajectory of the droplet center for several values of the initial offset $\Delta y$. For vanishing initial offset, $\Delta y = 0$, the droplet swims in a straight line to the left, while in the case $\Delta y \neq 0$ the droplet clearly moves away from the laser beam. As the droplet leaves the laser beam at $y/R = -1$, it continues to swim in a straight line until the surface is completely covered with *trans* surfactants and the droplet halts. Thus the swimming of pushed droplets is unstable with respect to an offset $\Delta y$ of the pushing laser beam.

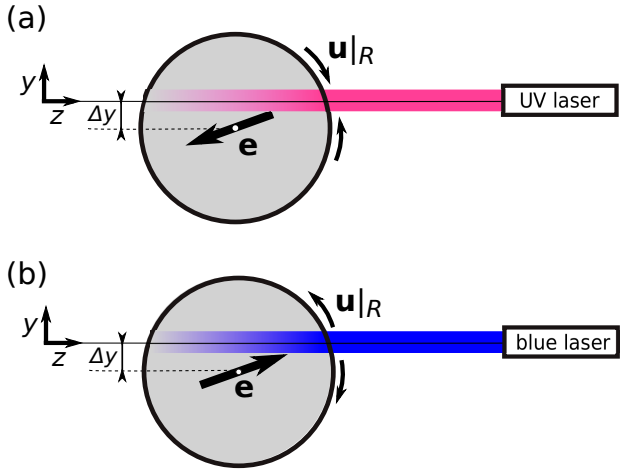

FIG. 7. (a) UV laser light is aimed at a spot which is offset by $\Delta y$ from the droplet center. The resulting Marangoni flow drives the droplet out of the laser beam. (b) Blue laser light is aimed at a spot which is offset by $\Delta y$ from the droplet center. The resulting Marangoni flow pulls the droplet back into the laser beam.

Finally, we discuss how the trajectories are influenced by the reduced relaxation rate $\kappa$, with which the surfactant mixture relaxes towards the equilibrium value $\phi_{\text{eq}}$. In Fig. 8 (a) we also plot trajectories for $\kappa = 2$ and 10 in addition to the default case $\kappa = 1$ for the same initial offset $\Delta y = -0.2R$. In all three cases the trajectories lie on top of each other, but for increasing $\kappa$ the droplet stops earlier. This is clear since the surfactant mixture relaxes faster to its equilibrium value, after the droplet has left the laser beam. Again, changing Marangoni number $M$ does not alter the results significantly.

## B. Pulling an absorbing droplet with blue light

In the following we present an alternative mechanism to drive an oil droplet by light. Here, the droplet of constant radius $R$ initially is in equilibrium with a water phase laden by *cis* surfactant. A blue laser beam with cross-sectional radius $\rho < R$ is focused on the center of the droplet and locally transforms the surfactant into the *trans* state [see Fig. 5 (b)], thereby lowering the surface tension of this region. The resulting Marangoni flow at the interface points away from the laser beam and thus pulls the droplet towards the laser beam. The advective current moves surfactants away from the laser beam, which are replenished by *cis* surfactants from the water phase. Again, the droplet oil phase completely ab-sorbs the incident light beam. In Sec. VI D we consider a transparent droplet, which is pulled by blue light.

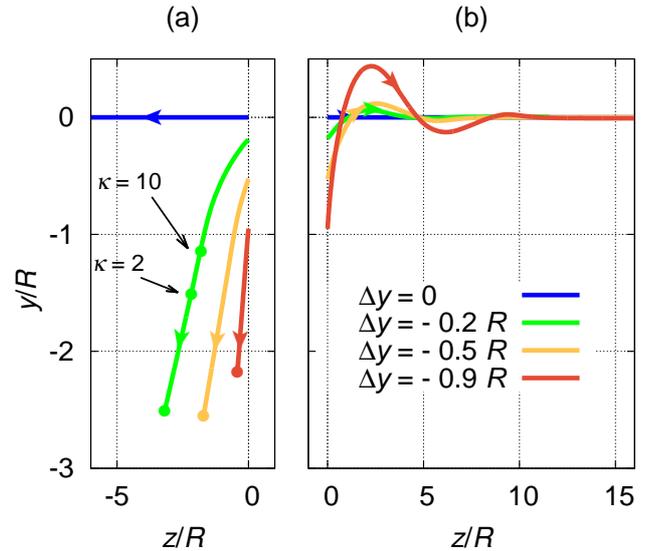

FIG. 8. (a) Trajectories of a droplet which is pushed by UV light. The droplet initially starts at $z = 0$ and $y = \Delta y$ and stops at the positions marked by dots. The laser is positioned at $y = 0$ and shines from right to left [compare Fig. 7 (a)]. Parameters are set to $M = 1$ and $\kappa = 1$, unless otherwise noted. The trajectories are symmetric w.r.t. changing the sign of $\Delta y$. (b) Trajectories of a droplet which is pulled by blue light. The droplet initially starts at $z = 0$ and $y = \Delta y$. The laser is positioned at $y = 0$ and shines from right to left [compare Fig. 7 (b)]. Again $M = \kappa = 1$.

### 1. Stationary solution of pulled droplet

For the numerical solution of Eq. (38) the order parameter $\phi$ is initially set to $\phi = \phi_{\text{eq}} = 1$. The blue laser beam with its circular patch of radius $\rho = 0.2R$ is implemented by setting $\phi = -1$ in the area of exposure. Figure 6, case (b) shows the stationary order parameter profile $\phi$ as well as the interfacial Marangoni flow $u_\theta|_R$. Since the maximum of $u_\theta|_R$ is at the front of the droplet, the droplet is a puller with $\beta = 1.4$. Note the different shape of $u_\theta|_R$ compared to the pushed droplet. The difference is due to the positive curvature $\sigma''(\phi) > 0$ of the nonlinear equation of state (37). Again, for increasing $\kappa$, the step in in the order parameter profile $\phi$ becomes steeper.

### 2. Pulling the droplet back to center

In Sec. VI A 3 we demonstrated the unstable swimming of the pushed droplet. The droplet pulled by the blue laser beam shows the opposite behavior. As sketched in Fig. 7 (b), the droplet with offset $\Delta y$ is pulled into the laser beam. This decreases the offset and the orientation vector $\mathbf{e}$ tilts towards and finally aligns along the laser beam. Figure 8(b) shows droplet trajectories for several initial offsets $\Delta y$. For $\Delta y = 0$ the droplet swims in a straight line to the right, while in the case $\Delta y \neq 0$ the droplet position relaxes towards $y = 0$ while performing damped oscillations about the stable swimming direction. Thus, the straight swimming trajectory along the laser beam is stable with respect to lateral excursions.

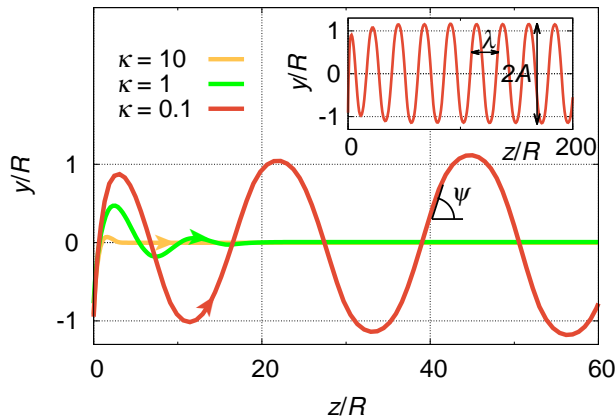

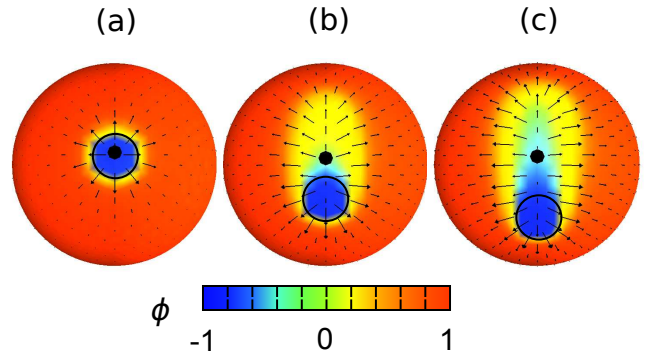

FIG. 9. Trajectories of a droplet, which is pulled by blue light for different relaxation rates $\kappa$. Initially, the droplet is placed at $z = 0$ and $y = -0.9R$. The laser is positioned at $y = 0$ and shines from right to left [compare Fig. 7 (b)]. Inset: For $\kappa = 0.1$ a stable oscillation with wavelength $\lambda$ and amplitude $A$ develops.

FIG. 10. Snapshots of the order parameter profile $\phi$ and the flow field $\mathbf{u}|_R$ at the interface of the pulled droplet for (a) $\kappa = 10$, (b) $\kappa = 1$, and (c) $\kappa = 0.1$. The area illuminated by the laser beam is shown by a circle and the bold dot indicates the swimming direction $\mathbf{e}$. The snapshots are taken from a supplemental movie.

Now, we discuss how the pulled droplet trajectories depend on $\kappa$. Figure 9 depicts them for an initial offset of $\Delta y = -0.9R$. For large relaxation rates such as $\kappa = 10$ (yellow curve in Fig. 9), the surfactants relax back to the *cis* conformation as soon as the illuminated region moves out of the laser beam. Hence, the swimming direction $\mathbf{e}$ is always directed towards the illuminated spot [see snapshot (a) in Fig. 10] and relaxes towards the beam direction as illustrated in the supplemental movie. However, a closer inspection of the yellow curve in Fig. 9 shows that the droplet crosses the $z$ axis before relaxing towards $y = 0$. This happens since the surfactant relaxation is not infinitely fast. The effect becomes even clearer for $\kappa = 1$ green curve in Fig. 9, where the lateral droplet position performs a damped oscillatory motion about the laser beam axis. Since the surfactant relaxation ($\kappa = 1$) is sufficiently slow compared to the droplet speed $M = 1$, the swimming direction $\mathbf{e}$ does not point towards the illuminated spot at early times [see snapshot (b) in Fig. 10]. The droplet crosses several times the $z$ axis before its direction aligns along the laser beam, as the supplemental movie shows. The movie also demonstrates how the step in the order parameter profile $\phi$ becomes steeper with increasing $\kappa$ when the stationary state is reached. This was already mentioned before. Interestingly, at very slow surfactant relaxation ($\kappa = 0.1$) the droplet performs a stable oscillatory motion about the beam axis, which is nicely illustrated by the supplemental movie. Increasing the Marangoni number $M$ increases swimming velocity and the oscillations occur at lower $\kappa^{-1}$.

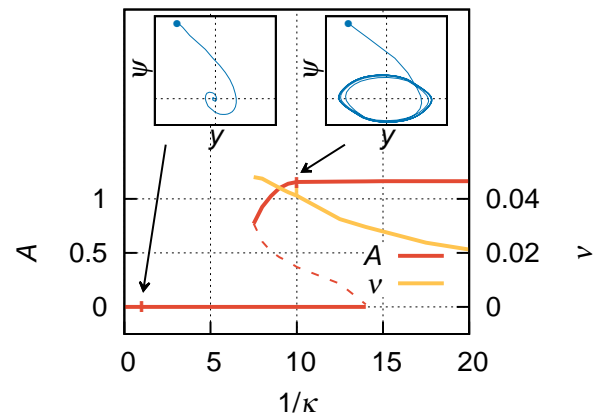

FIG. 11. Amplitude $A$ and wave number $\nu = 1/\lambda$ of the oscillating droplet trajectory (see inset of Fig. 9) plotted versus $\kappa^{-1}$. The two insets depict trajectories in $(y, \psi)$ phase space, where $\psi$ is the angle between droplet orientation and laser beam axis (see Fig. 9). Phase trajectories for $\kappa^{-1} = 1$ and 10 are shown and dots indicate the initial positions.

Figure 11 plots the amplitude $A$ of the stable oscillations versus $\kappa^{-1}$ and reveals a subcritical Hopf bifurcation. In the parameter range $\kappa^{-1} = 7.5$ to 14 both straight swimming (amplitude $A = 0$) and oscillatory motion ($A \neq 0$) occur depending on the initial lateral displacement $\Delta y$. Indeed, if $\Delta y$ is above the unstable branch of the Hopf bifurcation, plotted as dashed line in Fig. 11, the droplet assumes the oscillating state. The two swimming regimes are illustrated by phase portraits in orientation angle $\psi = \cos^{-1}(\mathbf{e} \cdot \mathbf{e}_z)$ versus $y$. They either reveal a stable fixpoint (inset for $\kappa^{-1} = 1$) or a stable limit cycle (inset for $\kappa^{-1} = 10$). Finally, we also plot the wave number $\nu = 1/\lambda$ of the oscillatory swimming motion along the $z$ axis. It decreases with $\kappa^{-1}$ since the droplet moves more persistently and thereby performs longer excursions from the beam axis.

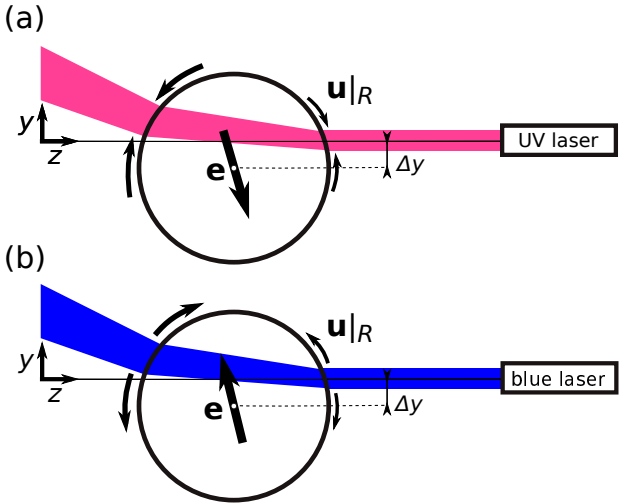

FIG. 12. (a) UV laser light is aimed at a transparent water droplet in oil, which is offset by $\Delta y$ from the laser beam. The resulting Marangoni flow at the two spots drives the droplet out of the beam. (b) Blue laser light is aimed at a water droplet in oil, which is offset by $\Delta y$ from the laser beam. The resulting Marangoni flow at the two spots pulls the droplet back into the laser beam.

In experiments, $\kappa = \tau/\tau_{eq}$ can again be tuned by adjusting the surfactant concentration $c_S$ in the bulk phase. We estimate the equilibration rate by $\tau_{eq}^{-1} = j \cdot 4\pi R^2$, where $j = k_a \cdot c_S$ is the flux of surfactants from the bulk to the droplet interface and $k_a$ is a typical adsorption rate constant. Using also $\tau = R^2/D_s$, one obtains $\kappa = k_a \cdot c_S \cdot 4\pi R^4/D_s$. With typical values $R = 50\mu m$, $D_s = 10^{-5} cm^2/s$, and $k_a = 10^9 m/(mol \cdot s)$, one then finds $\kappa \approx 10^5 l/mol \cdot c_S$, [72]. Thus, in order to observe the Hopf bifurcation at $\kappa \approx 0.1$, one has to set up an emulsion with surfactant density $c_S \approx 10^{-3} mmol/l$. For smaller $c_S$ we expect oscillations and for larger $c_S$ damped motion.

Finally, we note that we observed the same qualitative behavior as in Figs. 8-11 for a linear diffusive current $\mathbf{j}_D = D\nabla_s\phi$ and a linear equation of state for the surface tension $\sigma$. Hence, the origin of the Hopf bifurcation lies clearly in the nonlinear advection term $M\phi\mathbf{u}|_R$ of Eq. (38).

### C. Pushing a transparent droplet with UV light

In the following we discuss the case of an emulsion droplet with negligible light absorbance. The laser beam crosses the droplet and also actuates it at a second spot as illustrated in Fig. 12(a). Here, we focus on a water droplet immersed in a transparent oil phase laden with *trans* surfactants. But we will also comment on the in-verse case of an oil droplet in water. Due to the different refractive indices of oil and water, the transmitted beam is refracted at each interface according to the refraction law $n \sin\alpha = \hat{n} \sin\hat{\alpha}$. Here $\alpha$ and $\hat{\alpha}$ are the respective angles of the beam with respect to the surface normal in the oil and water phase while $n$ and $\hat{n}$ are the respective refraction indices. We apply the refraction law to partial beams of the incident light so that it widens while crossing and leaving the droplet. In what follows, we use $n = 1.45$ and $\hat{n} = 1.35$. Note that we neglect any reflection except for total reflection above the critical angle $\alpha_{max} = \arcsin(\hat{n}/n)$. For the emulsion droplet this implies that laser light is completely reflected if it hits the interface with a lateral distance to droplet center above $\Delta y_{max}/R = \hat{n}/n \approx 0.93$. The general mechanism for the light-induced Marangoni flow is the same as in Sec. VI A. It is directed away from each illuminated spot. Since the spots are well separated from each other, the droplet velocity vector is a superposition of the vectors induced by each spot.

Again, we start with a laser beam which is aimed at the center of the droplet. Due to refraction, the transmitted beam widens and the second illuminated spot is slightly larger than the first one. Thus, the velocity vector induced by the second spot is also slightly larger and slowly pushes the droplet towards the laser beam. This effect is hardly visible in our simulations. However, as soon as we introduce an offset $\Delta y$, the widening of the laser beam becomes stronger. The resulting velocity vector with orientation $\mathbf{e}$ pushes the droplet further away from the laser beam and also against the beam direction, as illustrated in Fig. 12 (a). Ultimately, the droplet leaves the beam completely. Figure 13 shows trajectories for various initial offsets. In the cases $\Delta y = -0.2R$ and $-0.5R$, the droplet initially moves in negative $y$ and positive $z$ direction [see also Fig. 12 (a)]. Once the second laser spot has sufficiently decreased in size, since part of the beam is totally reflected, the droplet moves in negative $z$ direction. It leaves the beam and finally stops. Thus, in analogy to the findings of Sec. VI A, the droplet is pushed out of the beam.

For an oil droplet immersed in water, the transmitted beam becomes more narrower. The droplet is still pushed out of the beam but the motion along $z$ direction is reversed. The corresponding trajectories are similar to the ones in Fig. 13, albeit reflected about the $y$-axis. If droplet and surrounding phase have equal refractive indices, the motion out of the beam is exactly along the $y$-axis. In all cases, the interfacial flow field is concentrated at the back of the droplet and the droplet is a pusher.

### D. Pulling a transparent droplet with blue light

Now, we study the effect of a blue light beam aimed at a water droplet, which is suspended in an oil phase laden with *cis* surfactants [see Fig. 12 (b)]. In this case, the Marangoni flow is directed away from the illuminated spots. At zero offset, $\Delta y = 0$, the droplet slowly moves along the negative $z$ direction. Any offset $\Delta y \neq 0$ pulls the droplet back into the beam with the velocity vector slightly tilted towards $-\mathbf{e}_z$ [see Fig. 12 (b)]. As in Sec. VI B, we use coupling strength $\kappa$ to distinguish between different regimes of motion.

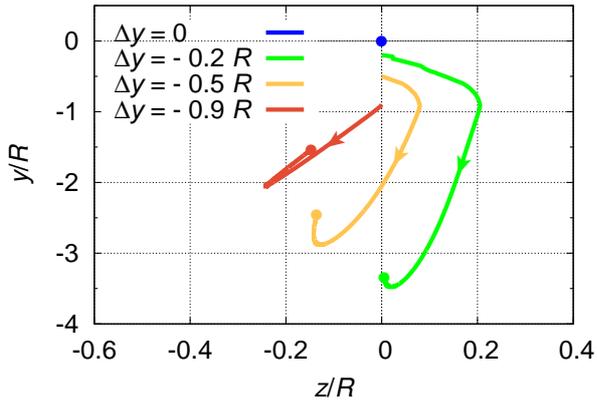

FIG. 13. Trajectories of a transparent droplet which is pushed by UV light. The droplet initially starts at $z = 0$ and $y = \Delta y$ and stops at the positions marked by dots. The laser is positioned at $y = 0$ and shines from right to left [compare Fig. 12 (a)]. Parameters are set to $M = 1$ and $\kappa = 1$. The trajectories are symmetric w.r.t. changing the sign of $\Delta y$.

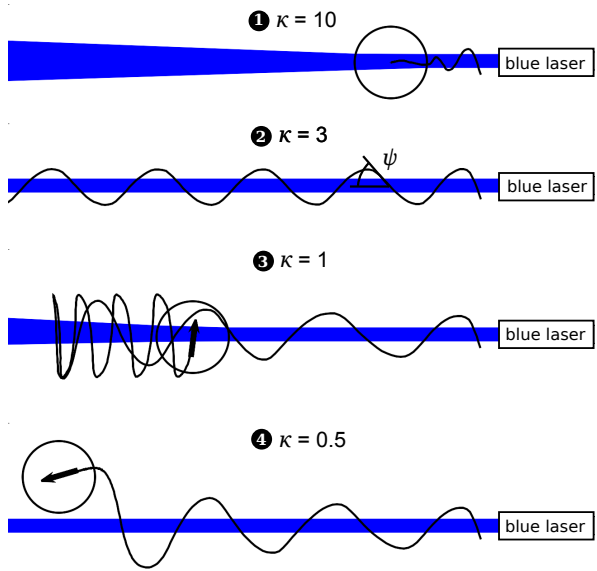

FIG. 14. Trajectories of a transparent water droplet suspended in oil, which is actuated by blue light for $\kappa = 10, 3, 1, 0.5$. The snapshots are taken from a supplemental movie. Trajectories of a transparent oil droplet suspended in water, which is actuated by blue light, are shown in a second supplemental movie. In all cases we set $M = 1$ and used an initial offset $\Delta y = -0.5$ [compare Fig. 12 (b)].

Figure 14 shows trajectories from a supplemental movie. In the case of strong coupling to the bulk phase, $\kappa = 10$, the droplet performs a damped oscillation about $y = 0$. The spatial resolution of our numerical method is not large enough to resolve the size difference between the two illuminated spots. Therefore, in the supplemental movie the droplet stops and does not move into the negative $z$ direction. Upon decreasing the the relaxation rate to values below $\kappa = 4.5$, the droplet undergoes a subcritical Hopf bifurcation and the droplet starts to oscillate about the laser beam. Figure 14 shows the trajectory from the supplemental movie; the droplet has already left the scene to the left. Figure 15 shows the subcritical bifurcation in the bottom graph, where amplitude $A$ and wave number $\nu$ are plotted versus $\kappa^{-1}$, or in the top phase portraits, where the limit cycle in case 2 is visible. Below $\kappa = 2.2$, the droplet changes its dynamics completely. After moving along the negative $z$ direction for a few droplet radii $R$, the droplet reverses its swimming direction and reaches a stationary oscillating state. The reversal occurs because the amplitude of the oscillation is so large that the size of the second spot decreases in size due to total reflection and the first spot pulls more strongly. This oscillation is characterized by larger amplitude $A$ and wave number $\lambda$ compared to the first oscillation state [see Fig. 15]. Finally, at relaxation rates below $\kappa = 0.54$, the droplet eventually leaves the beam and stops. For the experimental realization of different values of $\kappa$, we refer to the discussion in Sec. VI B 2.

For an oil droplet immersed in water, where total reflection does not occur, we only observe three states, in which the droplet moves against the laser beam: damped oscillations, stationary oscillations, and where the droplet ultimately leaves the laser beam. A supplemental movie illustrates the three cases.

## VII. CONCLUSIONS

A non-uniform surface tension profile $\sigma$ at the interface of an emulsion droplet generates flow fields at the interface and inside as well as outside of the droplet. The flow at the interface is directed along the gradient of $\sigma$. Using this Marangoni effect, the emulsion droplet becomes active. We decomposed the surface tension profile into spherical harmonics, $\sigma(\theta, \varphi) = \sum s_l^m Y_l^m$, and for this most general form of $\sigma$ we determined the full three-dimensional flow field inside $[\hat{\mathbf{u}}(\mathbf{r})]$, outside $[\mathbf{u}(\mathbf{r})]$, and at the interface $[\mathbf{u}|_R(\theta, \varphi)]$ of the droplet as a function of the expansion coefficients $s_l^m$. The swimming kinematics of the droplet follows from the droplet velocity vector $\mathbf{v}^D$, which solely depends on the coefficients $s_1^m$. The flow field outside of the droplet decays either as $1/r^3$ in the case of a neutral swimmer or as $1/r^2$ in the case of a pusher or a puller. The squirmer parameter $\beta$, for which we derived an expression in terms of the coefficients $s_1^m$ and $s_2^m$ for arbitrary swimming direction, enables to distinguish between these cases.

In the second part of this paper we presented two il-lustrative examples to demonstrate how gradients in the surface tension $\sigma$ can be achieved and studied the resulting droplet motion.

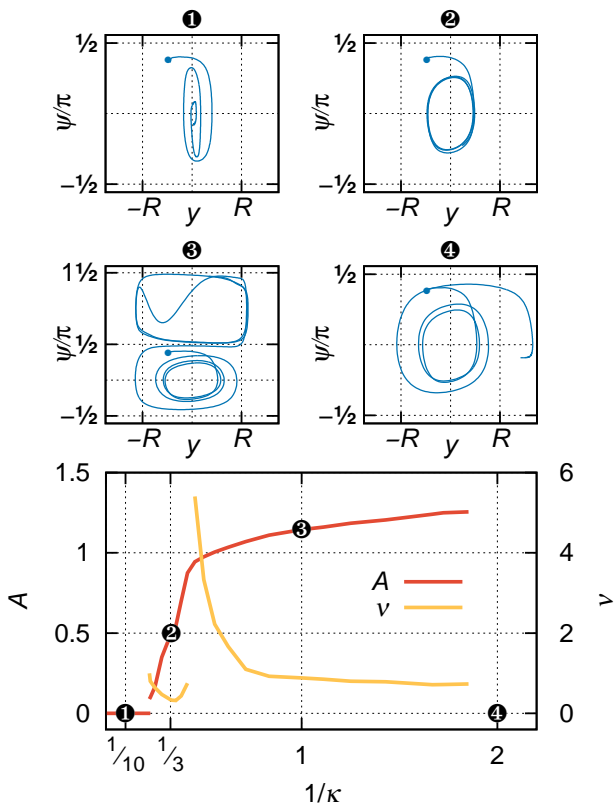

FIG. 15. Bottom: Amplitude $A$ and wave number $\nu = 1/\lambda$ of the oscillating droplet trajectories plotted versus $\kappa^{-1}$. Top: Trajectories in $(y, \psi)$ phase space at values of $\kappa^{-1}$ marked by numbers in the bottom plot. Here, $\psi$ is the angle between droplet orientation and laser beam axis, as indicated in Fig. 14. Dots indicate the initial positions. Fig. 14 shows the corresponding trajectories in $(z, y)$ space.

In the first example, we considered an initially surfactant free droplet, which adsorbes micelles formed by surfactants. The adsorbed micelle not only induces Marangoni flow in the proximity of the droplet interface but also radial fluid flow towards the adsorption site. The radial flow enhances the probability that other micelles adsorb at the same site. This mechanism leads to directed propulsion of an initially isotropic emulsion droplet if the micellar adsorption rate is sufficiently large. Clearly, the mechanism only works when surfactants are adsorbed through micelles. Single surfactants would not produce a sufficiently strong radial flow to spontaneously break the isotropic symmetry of the droplet. Our idealized example stresses the role which micelles play in generating directed motion in active emulsions. Therefore, it might contribute to understanding the self-propulsion of water and liquid-crystal droplets, which has been demonstrated in recent publications[25,26].

The second example considered a non-uniform mixture of two surfactant types in order to generate Marangoni flow. We used light-switchable surfactants based on the *trans-cis* isomerism of azobenzene to generate a non-uniform surfactant mixture. The analytic formulas for the flow field together with a diffusion-advection-reaction equation for the mixture order parameter determine the dynamics of the surfactant mixture and hence the droplet trajectory. We demonstrated that an emulsion droplet laden with *trans* surfactants, and either strongly adsorbing or transparent, can be pushed by a laser beam with UV light. However, the resulting straight trajectory is unstable with respect to displacing the droplet center relative to the laser beam axis. In contrast, a droplet laden with *cis* surfactants can be pulled into a laser beam with blue light. The straight trajectory is stable against lateral displacements. By decreasing the surfactant relaxation rate, the droplet develops an oscillatory trajectory about the laser beam via a subcritical Hopf bifurcation.

Having at hand analytic formulas for the full three-dimensional flow field, we are now able to fully discuss the recently introduced active emulsion droplet, where the surfactant mixture is generated by a bromination reaction[18,19]. We included thermal noise in the diffusion-advection-reaction equation of the mixture order parameter and currently study the coarsening dynamics of the surfactant mixture towards the stationary order parameter profile[71]. Thermal fluctuations in the composition of the surfactant mixture are responsible for the rotational diffusion of the swimming direction and thereby generate a persistent random walk of the active emulsion droplet.

## ACKNOWLEDGMENTS

We acknowledge financial support by the Deutsche Forschungsgemeinschaft in the framework of the collaborative research center SFB 910 and the research training group GRK 1558.

## Appendix A: Spherical harmonics

Throughout this paper we use the following definition of spherical harmonics:

$$Y_l^m(\theta, \varphi) = \sqrt{\frac{2l+1}{4\pi} \frac{(l-m)!}{(l+m)!}} \, P_l^m(\cos\theta) \, e^{im\varphi} \,,$$

with associated Legendre polynomials $P_l^m$ of degree $l$, order $m$, and with orthonormality:

$$\int_0^\pi \int_0^{2\pi} Y_l^m \, \overline{Y}_{l'}^{m'} \, \sin\theta \mathrm{d}\theta \mathrm{d}\varphi = \delta_{l,l'} \, \delta_{m,m'} \,,$$

where $\overline{Y}_l^m$ denotes the complex conjugate of $Y_l^m$.

**Appendix B: Fluid flow in the bulk**

Here we give the complete velocity field inside $\hat{\mathbf{u}} = \hat{\mathbf{v}} + \hat{\mathbf{w}}$ and outside $\mathbf{u} = \mathbf{v} + \mathbf{w}$ of the droplet in the droplet frame:

$$\hat{u}_r = \frac{-\eta}{2(\eta+\hat{\eta})}\left[\frac{r^2}{R^2}-1\right]\mathbf{e}_r \cdot \mathbf{v}^D + \frac{1}{\eta+\hat{\eta}}\sum_{l=1}^{\infty}\sum_{m=-l}^{l}\left[\frac{r^{l+1}}{R^{l+1}}-\frac{r^{l-1}}{R^{l-1}}\right]\frac{l(l+1)s_l^m}{4l+2}Y_l^m\,,\tag{B1a}$$

$$\hat{u}_\theta = \frac{-\eta}{2(\eta+\hat{\eta})}\left[\frac{2r^2}{R^2}-1\right]\mathbf{e}_\theta \cdot \mathbf{v}^D + \frac{1}{\eta+\hat{\eta}}\sum_{l=1}^{\infty}\sum_{m=-l}^{l}\left[(l+3)\frac{r^{l+1}}{R^{l+1}}-(l+1)\frac{r^{l-1}}{R^{l-1}}\right]\frac{s_l^m}{4l+2}\partial_\theta Y_l^m\,,\tag{B1b}$$

$$\hat{u}_\varphi = \frac{-\eta}{2(\eta+\hat{\eta})}\left[\frac{2r^2}{R^2}-1\right]\mathbf{e}_\varphi \cdot \mathbf{v}^D + \frac{1}{\eta+\hat{\eta}}\sum_{l=1}^{\infty}\sum_{m=-l}^{l}\left[(l+3)\frac{r^{l+1}}{R^{l+1}}-(l+1)\frac{r^{l-1}}{R^{l-1}}\right]\frac{ims_l^m}{4l+2}\frac{Y_l^m}{\sin\theta}\,,\tag{B1c}$$

$$u_r = \left(\frac{-\eta}{2(\eta+\hat{\eta})}\left[\frac{R}{r}-\frac{R^3}{r^3}\right]-\left[1-\frac{3R}{2r}+\frac{R^3}{2r^3}\right]\right)\mathbf{e}_r \cdot \mathbf{v}^D + \frac{1}{\eta+\hat{\eta}}\sum_{l=1}^{\infty}\sum_{m=-l}^{l}\left[\frac{R^l}{r^l}-\frac{R^{l+2}}{r^{l+2}}\right]\frac{l(l+1)s_l^m}{4l+2}Y_l^m\,,\tag{B1d}$$

$$u_\theta = \left(\frac{-\eta}{4(\eta+\hat{\eta})}\left[\frac{R}{r}+\frac{R^3}{r^3}\right]-\left[1-\frac{3R}{4r}-\frac{R^3}{4r^3}\right]\right)\mathbf{e}_\theta \cdot \mathbf{v}^D + \frac{1}{\eta+\hat{\eta}}\sum_{l=1}^{\infty}\sum_{m=-l}^{l}\left[(2-l)\frac{R^l}{r^l}+l\frac{R^{l+2}}{r^{l+2}}\right]\frac{s_l^m}{4l+2}\partial_\theta Y_l^m\,,\tag{B1e}$$

$$u_\varphi = \left(\frac{-\eta}{4(\eta+\hat{\eta})}\left[\frac{R}{r}+\frac{R^3}{r^3}\right]-\left[1-\frac{3R}{4r}-\frac{R^3}{4r^3}\right]\right)\mathbf{e}_\varphi \cdot \mathbf{v}^D + \frac{1}{\eta+\hat{\eta}}\sum_{l=1}^{\infty}\sum_{m=-l}^{l}\left[(2-l)\frac{R^l}{r^l}+l\frac{R^{l+2}}{r^{l+2}}\right]\frac{ims_l^m}{4l+2}\frac{Y_l^m}{\sin\theta}\,.\tag{B1f}$$

Note that for $r = R$, one recaptures Eqs. (17) and boundary condition $u_r = \hat{u}_r = 0$, while for $r \to \infty$: $\mathbf{u} = -\mathbf{v}^D$.

Note, by combining the outside field of the pumping solution $\mathbf{w}$ with the first term of the passive-droplet field $\mathbf{v}$, the stokeslet components, *i.e.*, terms with $\mathbf{u} \propto r^{-1}$ cancel each other:

$$u_r = -\mathbf{e}_r \cdot \mathbf{v}^D\left(1-\frac{R^3}{r^3}\right)$$
$$+\frac{1}{\eta+\hat{\eta}}\sum_{l=2}^{\infty}\sum_{m=-l}^{l}\left[\frac{R^l}{r^l}-\frac{R^{l+2}}{r^{l+2}}\right]\frac{l(l+1)s_l^m}{4l+2}Y_l^m\,,$$

$$u_\theta = -\mathbf{e}_\theta \cdot \mathbf{v}^D\left(1+\frac{R^3}{2r^3}\right)$$
$$+\frac{1}{\eta+\hat{\eta}}\sum_{l=2}^{\infty}\sum_{m=-l}^{l}\left[(2-l)\frac{R^l}{r^l}+l\frac{R^{l+2}}{r^{l+2}}\right]\frac{s_l^m}{4l+2}\partial_\theta Y_l^m\,,$$

$$u_\varphi = -\mathbf{e}_\varphi \cdot \mathbf{v}^D\left(1+\frac{R^3}{2r^3}\right)$$
$$+\frac{1}{\eta+\hat{\eta}}\sum_{l=2}^{\infty}\sum_{m=-l}^{l}\left[(2-l)\frac{R^l}{r^l}+l\frac{R^{l+2}}{r^{l+2}}\right]\frac{ims_l^m}{4l+2}\frac{Y_l^m}{\sin\theta}\,.$$

This shows that the droplet is a force-free swimmer[4]. Thus, in leading order the flow field is given by a stresslet with $\mathbf{u} \propto r^{-2}$. The squirmer parameter $\beta$ calculated in Sec. IV A determines the sign and the magnitude of the stresslet. In particular, if the coefficients $s_2^m$ vanish, the squirmer parameter also becomes zero ($\beta = 0$). Then, the flow field is less long-ranged and decays as $\mathbf{u} \propto r^{-3}$.

Finally, by adding the droplet velocity vector to our solution, one arrives at the velocity field $\mathbf{u}^L = \mathbf{u} + \mathbf{v}^D$ in the lab frame:

$$u_r^L = \frac{R^3}{r^3}\mathbf{e}_r \cdot \mathbf{v}^D$$
$$+\frac{1}{\eta+\hat{\eta}}\sum_{l=2}^{\infty}\sum_{m=-l}^{l}\left[\frac{R^l}{r^l}-\frac{R^{l+2}}{r^{l+2}}\right]\frac{l(l+1)s_l^m}{4l+2}Y_l^m\,,$$

$$u_\theta^L = \frac{-R^3}{2r^3}\mathbf{e}_\theta \cdot \mathbf{v}^D$$
$$+\frac{1}{\eta+\hat{\eta}}\sum_{l=2}^{\infty}\sum_{m=-l}^{l}\left[(2-l)\frac{R^l}{r^l}+l\frac{R^{l+2}}{r^{l+2}}\right]\frac{s_l^m}{4l+2}\partial_\theta Y_l^m\,,$$

$$u_\varphi^L = \frac{-R^3}{2r^3}\mathbf{e}_\varphi \cdot \mathbf{v}^D$$
$$+\frac{1}{\eta+\hat{\eta}}\sum_{l=2}^{\infty}\sum_{m=-l}^{l}\left[(2-l)\frac{R^l}{r^l}+l\frac{R^{l+2}}{r^{l+2}}\right]\frac{ims_l^m}{4l+2}\frac{Y_l^m}{\sin\theta}\,.$$

In this frame the velocity field satisfies the boundary condition $\mathbf{u}^L|_{r\to\infty} = 0$. Note that in this frame the radial component $u_r^L$ does not vanish at $r = R$.

## Appendix C: Lorentz reciprocal theorem

Applying the Lorentz reciprocal theorem to relate the flow fields of the pumping active droplet from Sect. II A and the passive droplet from Sect. II B to each other, one arrives at the alternative expression for the droplet velocity:[73]

$$\mathbf{v}^D = \frac{-1}{4\pi R^2} \frac{3\eta + 3\hat{\eta}}{2\eta + 3\hat{\eta}} \iint \mathbf{w}|_R \mathrm{d}A \ . \tag{C1}$$

Note that this generalizes the expression for rigid active spherical swimmers ($\hat{\eta} \to \infty$) in Ref.[59]. Using the surface flow field of the pumping droplet from Eqs. (12) in Eq. (C1), one obtains Eq. (21).

## Appendix D: Comparison with squirmer model

The presented solution $\mathbf{u}(\mathbf{r})$ for the flow field around an active droplet can be related to the axisymmetric squirmer model introduced by Lighthill[55] and later by Blake[56] as follows. The squirmer flow field can also be decomposed into a pumping active and a passive part, $\mathbf{u}^{sq} = \mathbf{w}^{sq} + \mathbf{v}^{sq}$, where $\mathbf{v}^{sq}$ is the usual Stokes flow field of a solid sphere, with a passive part, $\mathbf{v}^{sq}$. In order to match $\mathbf{w}$ with the known squirmer field $\mathbf{w}^{sq}$, one has to set

$$s_l = -(\eta + \hat{\eta}) \frac{4l + 2}{l(l+1)} \sqrt{\frac{4\pi}{2l+1}} B_l \ . \tag{D1}$$

This yields the correct flow field of a swimming squirmer with surface velocity field $u_\theta = \sum_{l=1}^{\infty} B_l V_l(\cos\theta)$, where $V_l = \frac{-2}{l(l+1)} P_l^1(\cos\theta)$. Here we used the notation of Blake[56].

## Appendix E: Written-out generalized squirmer parameter

The generalized squirmer parameter in Eq. (28) for a droplet swimming in an arbitrary direction can be written in terms of the coefficients $s_1^m$ and $s_2^m$ using Eqs. (22) and (24):

$$\beta = -\sqrt{\frac{27}{5}} \frac{\tilde{s}_2^0}{|\tilde{s}_1^0|} \ , \tag{E1a}$$

$$\tilde{s}_1^0 = \sqrt{(s_1^0)^2 - 2s_1^1 s_1^{-1}} \ , \tag{E1b}$$

$$\tilde{s}_2^0 = \left( \sqrt{6} \left[ s_2^2 (s_1^{-1})^2 + s_2^{-2} (s_1^1)^2 \right] - \sqrt{12} s_1^0 \left[ s_2^1 s_1^{-1} + s_2^{-1} s_1^1 \right] \right.$$
$$\left. + 2s_2^0 \left[ (s_1^0)^2 + s_1^1 s_1^{-1} \right] \right) / \left[ 2(s_1^0)^2 - 4s_1^1 s_1^{-1} \right] \ . \tag{E1c}$$

[1] A. Najafi and R. Golestanian, "Simple swimmer at low Reynolds number: Three linked spheres," Phys. Rev. E **69**, 062901 (2004).

[2] R. Dreyfus, J. Baudry, M. L. Roper, M. Fermigier, H. A. Stone, and J. Bibette, "Microscopic artificial swimmers," Nature **437**, 862 (2005).

[3] E. Gauger and H. Stark, "Numerical study of a microscopic artificial swimmer," Phys. Rev. E **74**, 021907 (2006).

[4] E. Lauga and T. R. Powers, "The hydrodynamics of swimming microorganisms," Rep. Prog. Phys. **72**, 096601 (2009).

[5] A. Walther and A. H. Müller, "Janus particles," Soft Matter **4**, 663 (2008).

[6] R. Golestanian, T. B. Liverpool, and A. Ajdari, "Propulsion of a Molecular Machine by Asymmetric Distribution of Reaction Products," Phys. Rev. Lett. **94**, 220801 (2005).

[7] W. F. Paxton, A. Sen, and T. E. Mallouk, "Motility of Catalytic Nanoparticles through Self-Generated Forces," Chem. Eur. J. **11**, 6462 (2005).

[8] J. L. Moran and J. D. Posner, "Electrokinetic locomotion due to reaction-induced charge auto-electrophoresis," J. Fluid. Mech. **680**, 31 (2011).

[9] T. Bickel, A. Majee, and A. Würger, "Flow pattern in the vicinity of self-propelling hot Janus particles," Phys. Rev. E **88**, 012301 (2013).

[10] I. Buttinoni, G. Volpe, F. Kümmel, G. Volpe, and C. Bechinger, "Active Brownian motion tunable by light," J. Phys.: Condens. Matter **24**, 284129 (2012).

[11] M. C. Marchetti, J. F. Joanny, S. Ramaswamy, T. B. Liverpool, J. Prost, M. Rao, and R. A. Simha, "Hydrodynamics of soft active matter," Rev. Mod. Phys. **85**, 1143 (2013).

[12] T. Ishikawa and T. J. Pedley, "Coherent Structures in Monolayers of Swimming Particles," Phys. Rev. Lett. **100**, 088103 (2008).

[13] A. A. Evans, T. Ishikawa, T. Yamaguchi, and E. Lauga, "Orientational order in concentrated suspensions of spherical microswimmers," Phys. Fluids **23**, 111702 (2011).

[14] F. Alarcón and I. Pagonabarraga, "Spontaneous aggregation and global polar ordering in squirmer suspensions," J. Mol. Liq. **185**, 56 (2013).

[15] A. Zöttl and H. Stark, "Hydrodynamics Determines Collective Motion and Phase Behavior of Active Colloids in Quasi-Two-Dimensional Confinement," Phys. Rev. Lett. **112**, 118101 (2014).

[16] M. Hennes, K. Wolff, and H. Stark, "Self-Induced Polar Order of Active Brownian Particles in a Harmonic Trap," Phys. Rev. Lett. **112**, 238104 (2014).

[17] A. Bricard, J.-B. Caussin, N. Desreumaux, O. Dauchot, and D. Bartolo, "Emergence of macroscopic directed motion in populations of motile colloids," Nature **503**, 95 (2013).

[18] S. Thutupalli, R. Seemann, and S. Herminghaus, "Swarming behavior of simple model squirmers," New J. Phys. **13**, 073021 (2011).

[19] M. Schmitt and H. Stark, "Swimming active droplet: A theoretical analysis," Europhys. Lett. **101**, 44008 (2013).

[20] S.-Y. Teh, R. Lin, L.-H. Hung, and A. P. Lee, "Droplet microfluidics," Lab Chip **8**, 198 (2008).

[21] R. Seemann, M. Brinkmann, T. Pfohl, and S. Herminghaus, "Droplet based microfluidics," Rep. Prog. Phys. **75**, 016601 (2012).

[22] M. M. Hanczyc, T. Toyota, T. Ikegami, N. Packard, and T. Sugawara, "Fatty acid chemistry at the oil-water interface: self-propelled oil droplets." J. Am. Chem. Soc. **129**, 9386 (2007).

[23] T. Toyota, N. Maru, M. M. Hanczyc, T. Ikegami, and T. Sugawara, "Self-Propelled Oil Droplets Consuming âĂIJFuelâĂİ Surfactant," J. Am. Chem. Soc. **131**, 5012 (2009).

[24] T. Banno, R. Kuroha, and T. Toyota, "pH-Sensitive Self-Propelled Motion of Oil Droplets in the Presence of Cationic Surfactants Containing Hydrolyzable Ester Linkages," Langmuir **28**, 1190 (2012).

[25] S. Herminghaus, C. C. Maass, C. Krüger, S. Thutupalli, L. Goehring, and C. Bahr, "Interfacial mechanisms in active emulsions," Soft Matter **10**, 7008 (2014).

[26] Z. Izri, M. N. van der Linden, S. Michelin, and O. Dauchot, "Self-Propulsion of Pure Water Droplets by Spontaneous Marangoni-Stress-Driven Motion," Phys. Rev. Lett. **113**, 248302 (2014).

[27]H. Kitahata, N. Yoshinaga, K. H. Nagai, and Y. Sumino, "Spontaneous motion of a droplet coupled with a chemical wave," Phys. Rev. E **84**, 015101 (2011).

[28]E. Tjhung, D. Marenduzzo, and M. E. Cates, "Spontaneous symmetry breaking in active droplets provides a generic route to motility," Proc. Natl. Acad. Sci. U. S. A. **109**, 12381 (2012).

[29]N. Yoshinaga, "Spontaneous motion and deformation of a self-propelled droplet," Phys. Rev. E **89**, 012913 (2014).

[30]S. Yabunaka, T. Ohta, and N. Yoshinaga, "Self-propelled motion of a fluid droplet under chemical reaction," J. Chem. Phys. **136**, 074904 (2012).

[31]A. Y. Rednikov, Y. S. Ryazantsev, and M. G. Velarde, "Drop motion with surfactant transfer in a homogeneous surrounding," Phys. Fluids **6**, 451 (1994).

[32]A. Y. Rednikov, Y. S. Ryazantsev, and M. G. Velarde, "Active Drops and Drop Motions due to Nonequilibrium Phenomena," J. Non-Equil. Thermody. **19**, 95 (1994).

[33]M. G. Velarde, A. Y. Rednikov, and Y. S. Ryazantsev, "Drop motions and interfacial instability," J. Phys.: Condens. Matter **8**, 9233 (1996).

[34]M. G. Velarde, "Drops, liquid layers and the Marangoni effect," Phil. Trans. R. Soc. Lond. A **356**, 829 (1998).

[35]N. Yoshinaga, K. H. Nagai, Y. Sumino, and H. Kitahata, "Drift instability in the motion of a fluid droplet with a chemically reactive surface driven by Marangoni flow," Phys. Rev. E **86**, 016108 (2012).

[36]K. Furtado, C. M. Pooley, and J. M. Yeomans, "Lattice Boltzmann study of convective drop motion driven by nonlinear chemical kinetics," Phys. Rev. E **78**, 046308 (2008).

[37]L. E. Scriven and C. V. Sternling, "The Marangoni Effects," Nature **187**, 186 (1960).

[38]M. D. Levan and J. Newman, "The effect of surfactant on the terminal and interfacial velocities of a bubble or drop," AIChE Journal **22**, 695 (1976).

[39]D. P. Mason and G. M. Moremedi, "Effects of non-uniform interfacial tension in small Reynolds number flow past a spherical liquid drop," Pramana **77**, 493 (2011).

[40]J. Blawzdziewicz, P. Vlahovska, and M. Loewenberg, "Rheology of a dilute emulsion of surfactant-covered spherical drops," Physica A **276**, 50 (2000).

[41]J. A. Hanna and P. M. Vlahovska, "Surfactant-induced migration of a spherical drop in Stokes flow," Phys. Fluids **22**, 013102 (2010).

[42]J. T. Schwalbe, F. R. Phelan Jr, P. M. Vlahovska, and S. D. Hudson, "Interfacial effects on droplet dynamics in Poiseuille flow," Soft Matter **7**, 7797 (2011).

[43]T. Sakai, "Surfactant-free emulsions," Curr. Opin. Colloid In. **13**, 228 (2008).

[44]A. Diguet, R.-M. Guillermic, N. Magome, A. Saint-Jalmes, Y. Chen, K. Yoshikawa, and D. Baigl, "Photomanipulation of a Droplet by the Chromocapillary Effect," Angew. Chem. Int. Edit. **121**, 9445 (2009).

[45]O. Pak and E. Lauga, "Generalized squirming motion of a sphere," J. Eng. Math. **88**, 1 (2014).

[46]S. Sadhal, P. Ayyaswamy, and J. Chung, *Transport Phenomena with Drops and Bubbles*, Mechanical Engineering Series, Springer, 1996.

[47]H. Lamb, *Hydrodynamics*, Cambridge University Press, 1932.

[48]J. Happel and H. Brenner, *Low Reynolds Number Hydrodynamics: With Special Applications to Particulate Media*, Mechanics of Fluids and Transport Processes, Springer Netherlands, 1983.

[49]R. Clift, J. R. Grace, and M. E. Weber, *Bubbles, Drops, and Particles*, Academic Press, 1978.

[50]V. A. Nepomniashchii, M. G. Velarde, and P. Colinet, *Interfacial Phenomena and Convection*, Chapman & Hall / CRC, 1 edition, 2002.

[51]O. S. Pak, J. Feng, and H. A. Stone, "Viscous Marangoni migration of a drop in a Poiseuille flow at low surface Péclet numbers," J. Fluid. Mech. **753**, 535 (2014).

[52]G. Batchelor, "The stress system in a suspension of force-free particles," J. Fluid. Mech. **41**, 545 (1970).

[53]T. Ishikawa, M. P. Simmonds, and T. J. Pedley, "Hydrodynamic interaction of two swimming model micro-organisms," J. Fluid. Mech. **568**, 119 (2006).

[54]S. Kim and S. Karrila, *Microhydrodynamics: Principles and Selected Applications*, Butterworth - Heinemann Series in Chemical Engineering, Dover Publications, 2005.

[55]M. J. Lighthill, "On the squirming motion of nearly spherical deformable bodies through liquids at very small reynolds numbers," Commun. Pur. Appl. Math. **5**, 109 (1952).

[56]J. R. Blake, "A spherical envelope approach to ciliary propulsion," J. Fluid. Mech. **46**, 199 (1971).

[57]M. T. Downton and H. Stark, "Simulation of a model microswimmer," J. Phys.: Condens. Matter **21**, 204101 (2009).

[58]L. Zhu, E. Lauga, and L. Brandt, "Self-propulsion in viscoelastic fluids: Pushers vs. pullers," Phys. Fluids **24**, 051902 (2012).

[59]H. A. Stone and A. D. Samuel, "Propulsion of microorganisms by surface distortions," Phys. Rev. Lett. **77**, 4102 (1996).

[60]I. Griffiths, C. Bain, C. Breward, D. Colegate, P. Howell, and S. Waters, "On the predictions and limitations of the Becker-Döring model for reaction kinetics in micellar surfactant solutions," J. Colloid Interf. Sci. **360**, 662 (2011).

[61]V. G. Levich, *Physicochemical Hydrodynamics*, Prentice-Hall, 1962.

[62]Here, we decompose the surface tension up to $l_{max} = 60$, which has proven to be sufficient to model an adsorbed micelle without leading to numerical problems such as the Gibbs phenomenon. More details about the simulation method can be found in Ref.[71].

[63]N. Van Kampen, *Stochastic Processes in Physics and Chemistry*, North-Holland Personal Library, Elsevier Science, 1992.

[64]W. Al-Soufi, L. Piñeiro, and M. Novo, "A model for monomer and micellar concentrations in surfactant solutions: Application to conductivity, NMR, diffusion, and surface tension data," J. Colloid Interf. Sci. **370**, 102 (2012).

[65]S. S. Dukhin, G. Kretzschmar, and R. Miller, *Dynamics of Adsorption at Liquid Interfaces: Theory, Experiment, Application*, volume 1, Elsevier, 1995.

[66]O. Karthaus, M. Shimomura, M. Hioki, R. Tahara, and H. Nakamura, "Reversible Photomorphism in Surface Monolayers," J. Am. Chem. Soc. **118**, 9174 (1996).

[67]J. Y. Shin and N. L. Abbott, "Using light to control dynamic surface tensions of aqueous solutions of water soluble surfactants," Langmuir **15**, 4404 (1999).

[68]K. Ichimura, S.-K. Oh, and M. Nakagawa, "Light-driven motion of liquids on a photoresponsive surface," Science **288**, 1624 (2000).

[69]J. Eastoe and A. Vesperinas, "Self-assembly of light-sensitive surfactants," Soft Matter **1**, 338 (2005).

[70]T. Krol, A. Stelmaszewski, and W. Freda, "Variability in the optical properties of a crude oil-seawater emulsion," Oceanologia **48**, 203 (2006).

[71]M. Schmitt and H. Stark, "Marangoni phase separation of surfactants drives active Brownian motion of emulsion droplet," In preparation (2015).

[72]B. Binks and D. Furlong, *Modern Characterization Methods of Surfactant Systems*, volume 83, CRC Press, 1999.

[73]R. S. Subramanian, "The Stokes force on a droplet in an unbounded fluid medium due to capillary effects," J. Fluid. Mech. **153**, 389 (1985).